\documentclass[journal,final,twoside]{IEEEtranTCOM}

\usepackage{cite}
\usepackage{amsmath}
\usepackage{amssymb}
\usepackage{amsthm}
\usepackage{thmtools}
\usepackage{graphicx}
\usepackage{color}
\usepackage{multicol}
\usepackage{multirow}

\theoremstyle{definition}

\newtheorem{remark}{Remark}
\declaretheorem[style=remark,qed=$\blacktriangleleft$]{example}

\DeclareMathOperator{\coeff}{coeff}
\DeclareMathOperator{\wt}{wt}

\DeclareMathOperator{\erfc}{erfc}

\allowdisplaybreaks

\title{Performance-Complexity-Latency Trade-offs of Concatenated RS-BCH Codes}
\author{Alvin Y. Sukmadji, \emph{Graduate Student Member, IEEE} and Frank R. Kschischang, \emph{Fellow, IEEE}\thanks{Submitted for
publication on April 20th, 2023. First revision submitted on November 1st, 2023.
Second revision submitted on February 16, 2024.
The authors are with the Edward S.\ Rogers Sr.\ Department
of Electrical \& Computer Engineering, University of Toronto, Canada.
Emails: \texttt{alvin.sukmadji@mail.utoronto.ca},
\texttt{frank@ece.utoronto.ca}.}}

\begin{document}
\maketitle
\begin{abstract}
Using a generating function approach, a computationally tractable
expression is derived to predict the frame error rate arising at the
output of the binary symmetric channel when a number of outer
Reed--Solomon codes are concatenated with a number of inner
Bose--Ray-Chaudhuri--Hocquenghem codes, thereby obviating the need for
time-consuming Monte Carlo simulations. Measuring (a) code performance
via the gap to the Shannon limit, (b) decoding complexity via an
estimate of the number of operations per decoded bit, and (c)
decoding latency by the overall frame length, a code search is performed
to determine the Pareto frontier for performance-complexity-latency
trade-offs.
\end{abstract}


\section{Introduction}
\label{sec:intro}

\IEEEPARstart{M}{otivated} by applications in high throughput optical communication
systems, this work examines the performance-complexity-latency
trade-offs of concatenated outer Reed--Solomon (RS) and inner
Bose--Ray-Chaudhuri--Hocquenghem (BCH) codes.  RS and BCH codes are
classical algebraic codes with efficient decoding algorithms, making
them attractive for systems, such as data center interconnects, that
demand low complexity and latency. In the Ethernet standard, for
example, an outer RS($544,514$) code, commonly known as KP4, is used
\cite{ieee8023,yang,bhoja,chagnon}.  As data rates increase, there is a
need to design stronger codes without significantly increasing
complexity or latency.  Concatenated codes \cite{forneycc} often
accomplish such needs.

Concatenated codes with RS outer codes have been widely deployed in
various applications~\cite{costello07}. For example, RS codes concatenated with
convolutional inner codes were implemented in the Voyager program
\cite{mceliece94,odenwalder,divsalar} and are also used in satellite communications
\cite{ccsds01,ccsds20,hershey}. Concatenated RS-RS codes are present in
the compact disc system \cite{schouhamerimmink}.  Concatenation with
random inner codes results in so-called Justesen codes
\cite{justesen72,elagooz}, which were the first example of a family of
codes having constant rate, constant relative distance, and constant
alphabet size.  Concatenated RS-BCH codes have also been proposed for
wireless communications \cite{yuan} and non-volatile memories
\cite{freudenberger}.

In optical communications, many recent forward error correction (FEC)
proposals involve concatenated codes.  These include systems where
hard-decision decoders are proposed for both outer and inner codes
\cite{tzimpargos,yang,liu,hu} or where soft-decision inner decoders are
used \cite{barakatain-lowldpcstc,barakatain-ldpczipper,nedelcu}.  In
so-called pseudo-product codes, the decoder iterates between inner and
outer codes \cite{g9751,wang,lee}.

Complexity and latency trade-offs for concatenated codes have been
studied previously in the context of LDPC-zipper codes
\cite{barakatain-ldpczipper,barakatain-higherordermodulation}.  Such
codes are of interest in long-haul optical communication, where a high
decoding latency is permissible. In contrast, the codes that we analyze
in this paper are those of interest in relatively low-latency
applications, such as data center interconnects. We are interested in
finding code parameters that provide best possible operating points in
the triadic trade-off space of performance, complexity, and latency.
Here performance is measured via the gap to the Shannon limit,
complexity is measured by estimating the worst-case number of
operations per decoded bit, and latency is measured by overall frame
length.  Various previous
papers have taken an analytical approach to evaluate the performance of
concatenated codes \cite{kasami,benedetto,lentner23,wang,ricciutelli}.
The authors of \cite{benedetto} introduce the concept
of a ``uniform interleaver,'' which allows for a simple derivation
(by averaging over a class of randomly chosen interleavers) of
various weight-enumerating functions associated with turbo codes.
We consider only a fixed interleaver and we use
a generating function, not to enumerate codeword weights, but rather to 
connect the bit-wise Hamming weight of an error pattern seen by the
BCH decoder to its
symbol-wise Hamming weight seen by the RS decoder. The previous literature
does not seem to have considered performance-complexity-latency trade-offs of
RS-BCH concatenated codes in the generality of this paper.

The rest of this paper is organized as follows. In
Sec.~\ref{sec:sysmodel}, we describe the concatenated RS-BCH system in
detail. Generating functions are used in
Secs.~\ref{sec:bitstobytes}--\ref{sec:predictingfer} to describe the
interaction between bit errors in the BCH codewords and symbol errors in
the RS codewords, giving rise to a computationally tractable formula
that accurately estimates the decoded frame error rate (FER),
eliminating the need for time-consuming Monte Carlo simulation. In
Sec.~\ref{sec:perfcomplattradeoffs}, we define metrics for performance,
complexity, and latency, and we present the results of a code search
establishing efficient operating points on the Pareto frontier in the
performance-complexity-latency trade-off space.

Throughout this paper, let $\mathbb{N}=\{0,1,2,\ldots\}$ be the set of
natural numbers. For any $n \in \mathbb{N}$, let $[n]=\{0,1,\ldots,n\}$
and let $[n]^*=[n]\setminus\{0\}$.  Let $\mathbb{F}_q$ denote the finite
field with $q$ elements.  A linear code $\mathcal{C}$ of block length
$n$, dimension $k$, and error-correcting radius $t$ is denoted
$\mathcal{C}(n,k,t)$.

\begin{figure*}[t]
\centering
\includegraphics{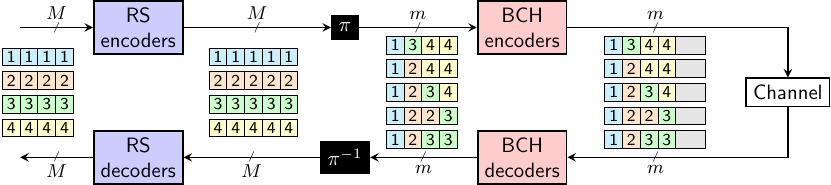}
\caption{Example of an RS-BCH concatenation with $M=4$ RS codewords and
$m=5$ BCH codewords.  Each square represents one ($B$-bit) RS symbol and
the number in each square represents the RS codeword index from which
that symbol originates.  Here, $K=4$, $N=5$, $k=4B$.}
\label{fig:system}
\end{figure*}

\section{System Model}
\label{sec:sysmodel}

Consider the concatenation of $M$ outer RS codes and $m$ inner BCH
codes, as illustrated in Fig.~\ref{fig:system}. For simplicity, we
assume throughout this paper that the outer codes are identical and the
inner codes are identical.  Each outer code is a (shortened) RS code
$\mathcal{C}_\text{RS}(N,K,T)$ over $\mathbb{F}_{2^B}$ for some positive
integer $B>1$.  We will sometimes refer to RS symbols as ($B$-bit)
\emph{bytes}.  Each inner code is a (shortened) BCH code
$\mathcal{C}_\text{BCH}(n,k,t)$ over $\mathbb{F}_2$, where $k$ is an
integer multiple of the byte size $B$.

We assume that all RS and BCH codewords are encoded systematically,
i.e., the first $K$ bytes in an RS codeword are information symbols and
the first $k$ bits in a BCH codeword are information bits. Furthermore,
we assume that interleaving---i.e., the mapping of symbols---between
outer and inner codes occurs in symbol-wise fashion: each RS symbol
(represented as a string of $B$ bits) is interleaved into the
information positions of a single BCH codeword.  Such symbol-wise
interleaving is commonly used in Ethernet standards and in optical
transport networks \cite{wang21,g9751}, and it simplifies our analysis;
see also Remark~\ref{remark:interleaving} in
Section~\ref{sec:predictingfer}.

Every symbol-wise interleaver induces an an $M\times m$ \emph{adjacency
matrix} $\mathbf{L}$, where $L_{i,j}$ (the entry in row $i$ and column
$j$ of $\mathbf{L}$) is equal to the number of RS symbols interleaved
from the $i$th outer code to the $j$th inner code.  Such an adjacency
matrix must satisfy the balance conditions
\begin{align}
\sum_{j=1}^m {L}_{i,j}=N\quad\text{and}\quad B\sum_{i=1}^M {L}_{i,j}=k,
\label{eq:sumofLs}
\end{align}
i.e., each outer code contributes $N$ symbols
and each inner code receives $k$ bits.

\begin{figure}[tbhp]
    \centering
    \includegraphics{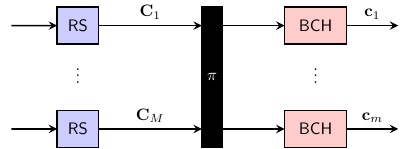}
    \caption{Concatenated RS-BCH encoder architecture.}
    \label{fig:encoder}
\end{figure}

Different symbol-wise interleavers may induce the same adjacency matrix
$\mathbf{L}$.  However, under bounded distance decoding of the
constituent codes at the output of a binary symmetric channel (which is
memoryless), each of them will have the same decoding performance. Thus
we define the following \emph{standard interleaver} that induces a given
$\mathbf{L}$.  For $i \in [M]^*$, let $\mathbf{C}_i \in
\mathbb{F}_{2^B}^N$ denote a codeword of the $i$th RS code.  We
subdivide $\mathbf{C}_i$ into $m$ consecutive \emph{strips}, writing
\[
\mathbf{C}_i=
\left(\mathbf{C}_{i,1}, \mathbf{C}_{i,2},\ldots, \mathbf{C}_{i,m}\right),
\]
where, for $j \in [m]^*$, the length of the $j$th strip is $L_{i,j}$
symbols.  Then, for $j \in [m]^*$, the $j$th BCH codeword is given as
\[
\mathbf{c}_j=
\left(\mathbf{c}_{1,j},\mathbf{c}_{2,j},\ldots,\mathbf{c}_{M,j},\mathbf{c}'_{j}\right),
\]
where $\mathbf{c}_{i,j}$ is equal to the strip $\mathbf{C}_{i,j}$
represented as a string of $BL_{i,j}$ bits, and $\mathbf{c}'_{j}$
denotes the $n-k$ parity bits of $\mathbf{c}_j$.  The encoder
architecture corresponding to this standard interleaver is summarized in
Fig.~\ref{fig:encoder}.

\begin{example}
In Fig.~\ref{fig:system}, we have $M=4$ RS codewords $\mathbf{C}_1,
\ldots, \mathbf{C}_4$ over $\mathbb{F}_{2^B}$ that are interleaved into
the information positions of the $m=5$ BCH codewords $\mathbf{c}_1,
\ldots, \mathbf{c}_5$.  Each RS codeword has block length $N=5$ and each
BCH codeword has dimension $k=4B$.
    
The RS codeword $\mathbf{C}_1$ contributes one symbol to each BCH
codeword, thus $L_{1,1}=L_{1,2}=L_{1,3}=L_{1,4}=L_{1,5}=1$.  The RS
codeword $\mathbf{C}_2$ contributes no symbols to $\mathbf{c}_1$, but it
does contribute one symbol to each of $\mathbf{c}_2$, $\mathbf{c}_3$,
and $\mathbf{c}_5$, and two symbols to $\mathbf{c}_4$; thus $L_{2,1}=0$,
$L_{2,2}=L_{2,3}=L_{2,5}=1$, and $L_{2,4} = 2$.  In similar fashion, we
can determine the other entries of the adjacency matrix, arriving at
\[
    \mathbf{L} = \left[ \begin{array}{ccccc}
        1&1&1&1&1\\
        0&1&1&2&1\\
        1&0&1&1&2\\
        2&2&1&0&0 \end{array}\right].
\]
Note that the sum of the entries in every row of $\mathbf{L}$ is $N=5$
and the sum of the entries in every column is $k/B=4$.
\end{example}

We are interested in estimating the decoded FER, i.e., the post-FEC FER.
The main challenge is to translate error rates across the BCH-RS
interface. In particular, BCH codes correct bit errors, while RS codes
correct byte errors. The latter can therefore correct more than one bit
for every corrected byte. The next two sections describe the
relationship between bit errors and byte errors and the derivation of an
expression of post-FEC FER assuming the binary symmetric channel.

\section{From Bits to Bytes}
\label{sec:bitstobytes}

\subsection{Hamming Weight}
Let $\mathbf{v}$ be a binary string of length $n$.  The (usual)
\emph{Hamming weight} of $\mathbf{v}$, denoted $\wt(\mathbf{v})$, is the
number of nonzero symbols that $\mathbf{v}$ contains.  Now, let $B$ be
any positive integer divisor of $n$. We define $w_B(\mathbf{v})$ as the
Hamming weight of $\mathbf{v}$ when $\mathbf{v}$ is regarded as an
element of $(\{0,1\}^B)^{n/B}$, i.e., as a string of length $n/B$ over
the alphabet $\{0,1\}^B$.  The zero element of $\{0,1\}^B$ is the
all-zero $B$-tuple $(0,\ldots,0)$, while all other elements of
$\{0,1\}^B$ are nonzero and thus contribute to $w_B(\mathbf{v})$ when
they occur in one of the $n/B$ positions of $\mathbf{v}$.

\begin{example}
The string $\mathbf{v}=000111\in\{0,1\}^6$ has Hamming weight
$
w_1(\mathbf{v}) = \wt(0,0,0,1,1,1) = 3.
$
When considered as a string over the quaternary ($B=2$) alphabet $\{ 00, 01,
10, 11 \}$, the same string has length $3$ and weight 
$
w_2(\mathbf{v}) =
\wt(00,01,11) = 2.
$
When considered as a string over the octonary ($B=3$) alphabet $\{ 000, 001,
\ldots, 111 \}$, the string has length $2$ and weight
$
w_3(\mathbf{v}) = \wt(000,111) = 1.
$
\end{example}

In the following, the elements of $\{0,1\}$ are referred to as \emph{bits},
while the symbols of $\{0,1\}^B$ are referred to as ($B$-bit) \emph{bytes}.

\subsection{Generating Functions}

To keep track of the relationships between the different notions of
Hamming weight, we will use generating functions.  For any positive
integer $n$, for any positive integer divisor $B$ of $n$, and for any
set $\mathcal{S} \subseteq \{ 0,1 \}^{n}$, let
\[
W_\mathcal{S}(x,y) = \sum_{\mathbf{v} \in \mathcal{S}}
x^{w_1(\mathbf{v})} y^{w_B(\mathbf{v})}.
\]
In $W_\mathcal{S}(x,y)$, the indeterminate $x$ tracks bit-wise Hamming
weight while the indeterminate $y$ tracks byte-wise Hamming weight.  For
$i,j \in \mathbb{N}$, the coefficient of the monomial $x^i y^j$ in
$W_\mathcal{S}(x,y)$, denoted as $\coeff_{x^i y^j}(W_\mathcal{S})$, is
equal to the number of vectors of $\mathcal{S}$ having bit-wise Hamming
weight $i$ while simultaneously having byte-wise Hamming weight $j$.

As usual, since Hamming weight is additive
over components, the key advantage of using generating functions
is that the Cartesian product relation
\begin{equation}
W_{\mathcal{S}^L}(x,y) = W_\mathcal{S}^L(x,y)
\label{eqn:cartesian}
\end{equation}
holds, where $\mathcal{S}^L$ denotes the $L$-fold Cartesian product of the
set $\mathcal{S}$ with itself.

Fix a byte size $B$, and let $W_{B,1}(x,y) = W_{\{ 0,1 \}^B }(x,y)$ be
the generating function corresponding to a single $B$-bit byte.  We then
have
\[
W_{B,1}(x,y) = 1 + ((1+x)^B-1)y = 1 + y \sum_{i=1}^B \binom{B}{i} x^i,
\]
which is explained as follows. The constant coefficient $\coeff_{x^0
y^0}(W_{B,1}) = 1$, indicating that exactly one word (the all-zero word)
simultaneously has bit-wise Hamming weight zero and byte-wise Hamming
weight zero.  For $i>0$, $\coeff_{x^i y} = \binom{B}{i}$, indicating
that there are $\binom{B}{i}$ non-zero bytes having bitwise Hamming
weight $i$ and byte-wise Hamming weight 1.  For $j > 1$, $\coeff_{x^i
y^j} = 0$, as there are no words with byte-wise Hamming weight greater
than 1.

More generally, let $W_{B,L}(x,y)$ denote the generating function for
$\{ 0, 1 \}^{BL}$, the collection of all binary strings of length $BL$,
or equivalently, strips of length $L$ ($B$-bit) bytes.  The Cartesian
product relation (\ref{eqn:cartesian}) then gives
\begin{align}
W_{B,L}(x,y) = W_{B,1}^L(x,y) = (1 + ((1+x)^B-1)y)^L \label{eqn:genfun}.
\end{align}
If $L=0$, then the strip $\mathbf{v}$ is empty. For the empty string
$\mathbf{v}$, we define $w_1(\mathbf{v}) = w_B(\mathbf{v}) = 0$ and
$W_{B,0}(x,y)=1$.

For our analysis of a concatenated RS-BCH code with $M$ outer RS codes,
we now extend the idea of enumerating the Hamming weights of a single
strip to enumerating the various Hamming weights of a string composed of
$M$ strips of bytes, with each strip possibly having a different length.
Let $\lambda = (L_1, \ldots, L_M)$ be an $M$-tuple of nonnegative
strip-lengths $L_1,\ldots, L_M$ and let $k = B(L_1 + \cdots + L_M)$.
Let $\mathbf{v}=(\mathbf{v}_1,\ldots,\mathbf{v}_M,\mathbf{v}') \in \{
0,1 \}^n$ be a binary string of length $n \geq k$, with strips
$\mathbf{v}_i \in \{ 0,1 \}^{B L_i}$ for $i \in [M]^*$, and
$\mathbf{v}'\in\{0,1\}^{n-k}$.  We say that $\mathbf{v}$ is partitioned
into strips according to $\lambda$.  In our application, $\mathbf{v}$ is
the $j$th BCH codeword, $\lambda$ is given by the $j$th column of the
adjacency matrix $\mathbf{L}$, the $M$ strips
$\mathbf{v}_1,\ldots,\mathbf{v}_M$ correspond to the outer RS symbols
interleaved to that codeword, and $\mathbf{v}'$ corresponds to the
parity bits of the BCH code (which are not passed to the outer codes).

For any subset $\mathcal{S} \subseteq \{ 0, 1 \}^n$ of binary strings
partitioned according to $\lambda$, let
\[
W_\mathcal{S}(x,y_1,\ldots,y_M) = \sum_{\mathbf{v} \in \mathcal{S}}
x^{w_1(\mathbf{v})}y_1^{w_B(\mathbf{v}_1)}y_2^{w_B(\mathbf{v}_2)}
\cdots y_M^{w_B(\mathbf{v}_M)}
\]
be a multivariate generating function.  The indeterminate $x$ tracks the
bit-wise Hamming weight of the elements of $\mathcal{S}$, while for
$i\in[M]^*$, $y_i$ keeps track of the byte-wise Hamming weights of the
strips $\mathbf{v}_i$ within the elements of $\mathcal{S}$.

Now, taking $\mathcal{S}= \{ 0,1 \}^n$ (the set of all binary
$n$-tuples), let
\[
W_{B,n,\lambda}(x,y_1,y_2,\ldots,y_M)= W_{ \{0,1\}^n }(x,y_1,\ldots,y_M)
\]
be the corresponding generating function.  Similar to before, since
Hamming weight is additive over components, we can represent
$W_{B,n,\lambda}$ as
\begin{align}
W_{B,n,\lambda}&(x,y_1,y_2,\ldots,y_M)\nonumber\\
&= W_{B,L_1}(x,y_1)\cdot\cdots\cdot W_{B,L_M}(x,y_M)\cdot(1+x)^{n-k} \nonumber\\
&= \left[\prod_{i=1}^M (1 + ((1+x)^B-1)y_i)^{L_i}\right](1+x)^{n-k}.
\label{eq:genfuncW}
\end{align}
The coefficient of $x^\ell y_1^{r_1}y_2^{r_2}\cdots y_M^{r_M}$ of
$W_{B,n,\lambda}$, denoted by $\coeff_{x^\ell y_1^{r_1}y_2^{r_2}\cdots
y_M^{r_M}}(W_{B,n,\lambda})$, represents the number of patterns in
$\{0,1\}^n$ of bit-wise Hamming weight $\ell$, having byte-wise Hamming
weights $r_1,r_2,\ldots,r_M$ in strips $1,2,\ldots,M$, respectively.

\begin{example}
Consider $B=2$, $M=2$, $\lambda=(1,1)$ (thus $k=2(1+1)=4$) and $n=5$. We
then have
\begin{multline*}
W_{B,n,\lambda}(x,y_1,y_2)\\
 = (1 + ((1+x)^2-1)y_1)(1 + ((1+x)^2-1)y_2)(1+x)\\
 = 1 + x + 2xy_1 + 2xy_2 + 3x^2y_1 + 3x^2y_2 + 4x^2y_1y_2 \\
 \quad + x^3y_1 + x^3y_2 + 8x^3y_1y_2 + 5x^4y_1y_2 + x^5y_1y_2.
\end{multline*}
Binary strings in $\{0,1\}^5$ having binary Hamming weight $\ell$, while
having byte-wise Hamming weight $r_1$ and $r_2$ in $c_1$ and $c_2$,
respectively, can be grouped as shown in Table~\ref{tab:patterns}.
\end{example}

\begin{table*}[htbp]
    \centering
    \caption{Length-5 Binary Strings with $B=2$, $M=2$, $\lambda=(1,1)$}
\renewcommand{\arraystretch}{1.1}
    \begin{tabular}{|c|c|c|c|c|l|}
    \hline
$\ell$ & $r_1$ & $r_2$ & \# Strings & Term & Strings \\\hline
$0$ & $0$ & $0$ & $1$ & $1$ & $(00\,00\,0)$ \\\hline
$1$ & $0$ & $0$ & $1$ & $x$ & $(00\,00\,1)$ \\\hline
$1$ & $1$ & $0$ & $2$ & $xy_1$ & $(10\,00\,0)$, $(01\,00\,0)$ \\\hline
$1$ & $0$ & $1$ & $2$ & $xy_2$ & $(00\,10\,0)$, $(00\,01\,0)$ \\\hline
$2$ & $1$ & $0$ & $3$ & $x^2y_1$ & $(11\,00\,0)$, $(10\,00\,1)$, $(01\,00\,1)$ \\\hline
$2$ & $0$ & $1$ & $3$ & $x^2y_2$ & $(00\,11\,0)$, $(00\,10\,1)$, $(00\,01\,1)$ \\\hline
$2$ & $1$ & $1$ & $4$ & $x^2y_1y_2$ & $(10\,10\,0)$, $(10\,01\,0)$, $(01\,01\,0)$, $(01\,10\,0)$ \\\hline
$3$ & $1$ & $0$ & $1$ & $x^3y_1$ & $(11\,00\,1)$ \\\hline
$3$ & $0$ & $1$ & $1$ & $x^3y_2$ & $(00\,11\,1)$ \\\hline
$3$ & $1$ & $1$ & $8$ & $x^3y_1y_2$ & $(11\,10\,0)$, $(11\,01\,0)$, $(10\,11\,0)$, $(01\,11\,0)$, $(10\,10\,1)$, $(10\,01\,1)$, $(01\,10\,1)$, $(01\,01\,1)$\\\hline
$4$ & $1$ & $1$ & $5$ & $x^4y_1y_2$ & $(11\,11\,0)$, $(11\,10\,1)$, $(11\,01\,1)$, $(10\,11\,1)$, $(01\,11\,1)$ \\\hline
$5$ & $1$ & $1$ & $1$ & $x^5y_1y_2$ & $(11\,11\,1)$ \\\hline
\end{tabular}
    \label{tab:patterns}
\end{table*}

\section{Predicting the Frame Error Rate}
\label{sec:predictingfer}

\subsection{Miscorrection-Free Analysis}

Suppose now that binary strings of length $n$ are drawn at random
according to any probability distribution satisfying the property that
strings $\mathbf{v}$ with a fixed bit-level Hamming weight
$w_1(\mathbf{v})$ all have the same probability of occurrence.  We refer
to any such distribution as being \emph{quasi-uniform by weight}.

For example, if the individual bits within $\mathbf{v}$ are drawn
independently according to a Bernoulli($p$) distribution (where $p$
denotes the probability of drawing ``1''), then strings $\mathbf{v}$
with $w_1(\mathbf{v}) = i$ all have the same probability of occurrence,
namely $p^i (1-p)^{n-i}$.

\begin{figure}[t]
    \centering
    \includegraphics{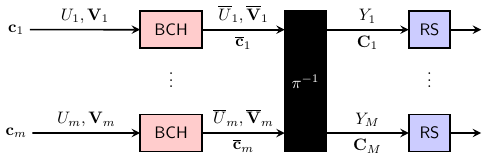}
    \caption{Concatenated RS-BCH decoder architecture.}
    \label{fig:decoderflow}
\end{figure}

We consider the decoder architecture summarized in
Fig.~\ref{fig:decoderflow}.  We begin by analyzing
\emph{miscorrection-free} decoders, i.e., hypothetical decoders that are
able to correct error patterns of weight up to their decoding radius,
while declaring a decoding failure when error patterns of higher weight
are encountered.  A miscorrection-free decoder never increases the
weight of an error pattern by decoding to an incorrect codeword, which
is an unrealistic assumption when the decoding radius is small.  The
influence of miscorrections is incorporated into the analysis in
Section~\ref{subsec:miscorrection}.

Suppose that all-zero codewords are transmitted.  A miscorrection-free binary
$t$-error correcting (BCH) decoder at the output of a binary symmetric channel
would map all strings (representing error patterns) of bit-level Hamming weight
$t$ or fewer to the all-zero string, while acting as an identity map on strings
of Hamming weight greater than $t$.  The output of this decoder would then also
have a distribution that is quasi-uniform by weight.

More concretely, let $U_j$ denote the number of bit errors
(i.e., the binary Hamming weight) in the
received BCH$(n,k,t)$ codeword $\mathbf{c}_j=(\mathbf{c}_{1,j},\ldots
\mathbf{c}_{M,j},\mathbf{c}_j')$.
The probability mass function of
$U_j$ depends on the type of channel that is used. For example, for a
binary symmetric channel with crossover probability
$p$, we have
\begin{align}
\Pr(U_{j}=\ell)=\binom{n}{\ell} p^{\ell}(1-p)^{n-\ell}\quad\text{for all}~ \ell \in [n].
    \label{eq:bit-error-prob}
\end{align}
Let $\overline{U}_j$ be the number of bit errors of the BCH codeword
$\overline{\mathbf{c}}_j=(\overline{\mathbf{c}}_{1,j},\ldots,
\overline{\mathbf{c}}_{M,j},\overline{\mathbf{c}}_j')$ remaining at the
output of the $j$-th BCH decoder, i.e., after inner decoding.  In the
case of a miscorrection-free inner decoder, we have
\begin{align}
    \Pr(\overline{U}_j=\ell)=\begin{cases}
        \sum_{i=0}^t \Pr(U_j=i) & \ell=0,\\
        0 & \ell=1,\ldots,t,\\
        \Pr(U_j=\ell) & \ell=t+1,\ldots,n.
    \end{cases}
    \label{eq:proboutputbch}
\end{align}

We can now compute the number of byte errors in strip
$\overline{\mathbf{c}}_{i,j}$ for each $i\in[M]^*,j\in[m]^*$.  Let
$\mathbf{V}_j=(V_{1,j},\ldots,V_{M,j})$, where $V_{i,j}$ is the number of byte errors
in the strip $\overline{\mathbf{c}}_{i,j}$. Then for all $\ell\in[n]$
and $\mathbf{v}_j\in [L_{1,j}]\times\cdots\times[L_{M,j}]$,
\begin{multline*}
    \Pr(\mathbf{\mathbf{V}}_j=\mathbf{v}_j) = \sum_{\ell=0}^{n}
    \Pr(\mathbf{\mathbf{V}}_j=\mathbf{v}_j\mid \overline{U}_{j}=\ell)\Pr(\overline{U}_{j}=\ell)\\
    =\sum_{\ell=0}^{n}\frac{\coeff_{x^\ell y_1^{v_{1,j}}\cdots y_M^{v_{M,j}}}
    (W_{B,n,\lambda_j})}{\binom{n}{\ell}}\Pr(\overline{U}_{j}=\ell),
\end{multline*}
where $W_{B,n,\lambda_j}$ is as defined in \eqref{eq:genfuncW} with
$\lambda_j=(L_{1,j},\ldots,L_{M,j})$.

Turning now to the perspective of the outer RS decoders, for any $i\in[M]^*$,
let $Y_i$ denote the number of byte errors passed to the $i$th RS codeword
$\mathbf{C}_i=(\mathbf{C}_{i,1},\ldots,\mathbf{C}_{i,m})$. Then
$Y_i=V_{i,1}+\cdots+V_{i,m}$. Written as a vector random
variable,
\begin{align}
    (Y_1,\ldots,Y_M) &= (V_{1,1},\ldots,V_{M,1})+\cdots+(V_{1,m},\ldots,V_{M,m})\nonumber\\
    &=\mathbf{V}_1+\cdots+\mathbf{V}_m.
    \label{eq:YM}
\end{align}
The $\mathbf{V}_j$'s are mutually independent, because they are obtained
from independent BCH decoders. The joint distribution of
$(Y_1,\ldots,Y_M)$ can therefore be obtained by convolving the
distributions of the $\mathbf{V}_j$'s.

An RS decoder is unable to correct an error pattern containing more than
$T$ bytes in error. A \emph{frame error} (FE) occurs if one or more RS
decoders are unable to correct their error patterns.  The frame error
rate is given by
\begin{align}
    \Pr(\text{FE}) = \Pr\left((Y_1> T)\vee\cdots\vee(Y_M> T)\right).
    \label{eq:fer}
\end{align}
Computing the exact frame error rate is difficult due to the correlation
between the symbols in different RS codewords.  However, we can bound
the frame error rate via the union bound
\begin{align}
    \Pr(\text{FE})\leq\sum_{i=1}^M \Pr(Y_i>T).
    \label{eq:unionbound}
\end{align}

\begin{figure}[t]
    \centering
    \includegraphics{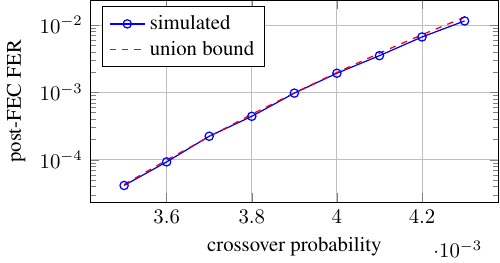}
    \caption{FER versus crossover probability for 8 $\times$ RS(544, 514, 15) outer code and 50 $\times$
    BCH(940, 880, 6) inner code. We assume that both outer and inner decoders are miscorrection-free.}
    \label{fig:misc}
\end{figure}

Fig.~\ref{fig:misc} compares the computed FER values obtained from
(\ref{eq:unionbound}) against those computed via a Monte Carlo
simulation assuming miscorrection-free decoders. We observe that the
simulated data points agree closely with those computed using
(\ref{eq:unionbound}).

\begin{remark}
While we have chosen to estimate the FER, our generating function
approach is easily modified to estimate the bit error rate (BER)
instead. See Appendix~\ref{sec:predictingser} for details.
\end{remark}
\begin{remark}
In the crossover probability regime of most interest to optical
communications, bit-error patterns typically result in one bit error per
byte error. Thus, we would expect that performing bit-level interleaving
rather than symbol-level interleaving (as analyzed here) would lead to
similar post-FEC FER or BER.
\label{remark:interleaving}
\end{remark}
\begin{remark}
Generalization of this analysis to the situation
where inner and outer codes have varying code parameters
is straightforward, though cumbersome. Multiple generating
functions would need to be introduced to track the error weights
transferred from the various inner codes to the outer codes.
The essential insight remains that the BCH decoders operate independently,
leading to a sum of independent random variables as in \eqref{eq:YM},
for the error weight seen by any particular RS decoder.
\end{remark}

\subsection{Effects of Miscorrection}
\label{subsec:miscorrection}

In practice, algebraic decoders will not be miscorrection-free.  A
miscorrection, i.e., correction to a codeword not actually transmitted,
can occur when the weight of the error pattern exceeds the decoding
radius. For simplicity, this section will consider miscorrections in the
inner BCH codes, but the result can be generalized to include
miscorrections in the outer RS codes.  We continue to assume that
all-zero codewords are transmitted.

In general, any decoding procedure can be thought as a transformation of
the error weight of the received word. More concretely, suppose that $U$
and $\overline{U}$ respectively denote the number of bit errors of the
vector at the input and output of a BCH decoder with decoding radius
$t$. Also, let $p_u=\Pr(U=u)$. In the case of a bounded-distance
algebraic decoder, miscorrections may occur when $u$ is higher than the
decoding radius.  Thus, for $u\geq t+1$,
\[
\Pr(\overline{U}=u)=f(p_{u-t},p_{u-t+1},\ldots,p_{u+t})
\]
for some function $f$ which depends on the code.

Determining the weight distribution of the output of an algebraic
decoder conditioned on the all-zero codeword being sent is, in general,
a difficult problem.  Error weights of $t$ or smaller are mapped to zero
error weight (since the decoder can correct the errors) while, depending
on the code, error weights larger than $t$ may result in an output error
weight that depends on the particular input pattern.  In case of a
decoding failure the output error weight will be equal to the input
error weight, while in the case of a miscorrection, the output error
weight will match the weight of a nonzero codeword.

To estimate the output error weight from the input error weight,
we will make the following assumptions.
\begin{enumerate}
\item When a miscorrection occurs, a word of error weight $w > t$ is
miscorrected to a codeword of weight $w+t$.
\item Given that the received word has more than $t$ errors, the conditional
miscorrection probability of a $t$-error correcting BCH code
of length $n$ shortened from length $n_{\text{BCH}}$,
is at most \cite{mceliece,sukmadji23}
\[
    \frac{1}{t!}\left(\frac{n}{n_{\text{BCH}}}\right)^t.
\]
This assumption allows us to ignore the effects
of miscorrections when $t$ is sufficiently large.
\end{enumerate}
The first assumption is justified as follows.  In the crossover
probability regime of interest ($p<10^{-3}$), the error weight $w$ of
the received word will only very rarely exceed $np+7\sqrt{np}$, i.e., 7
standard deviations above the mean, thus $w \ll n$.
Let $A_i$ denote the number of codewords of weight $i$
in a $t$-error-correcting binary BCH code of length $n$.
Clearly $A_i=0$ for all $i\in\{1,2,\ldots,2t\}$; and it
is known \cite{peterson} that
$A_i \approx 2^{-(n-k)}\binom{n}{i}$
for $i\in\{2t+1,\ldots,n-2t-1\}$.
Suppose that the all-zero codeword is sent and that
the decoder miscorrects a word of weight $w > t$ to a
codeword of weight $w'\in\{w-t,w-t+1,\ldots,w+t\}$ at random.
The probability that the codeword will have weight $w'$ would then be
\[
\frac{A_{w'}}{A_{w-t}+\cdots+A_{w+t}}.
\]
For $w \ll n$, we have $A_{w+t}\gg
A_{w+t-1}\gg\cdots\gg A_{w-t}$,
so $\Pr(w'=w+t) \approx 1$.
Thus, in the event of a miscorrection,
it is very likely that a word of weight $w$ is miscorrected to a
codeword of weight $w+t$ rather than to some other weight.

We now define a sequence of genie-aided decoders $D_{\ell}$,
for $\ell = 0, 1, 2, \ldots$, in which $D_{\ell}$ declares
a decoding failure upon receipt of a word of error weight $t+1+\ell$ or
greater, but otherwise attempts to decode.  Thus $D_0$ is the miscorrection-free
decoder, $D_1$ will suffer miscorrections due to attempting to
decode words of error weight $t+1$,
$D_2$ will suffer miscorrections due to attempting to
decode words of error weight $t+1$ and $t+2$, etc.

We can estimate the performance of
these decoders under the assumptions stated above.
Table~\ref{tab:bsc-misc} gives the post-FEC BER for length-500 BCH
codes with $t=1,2,3,4,5$ assuming a binary symmetric
channel with crossover probability
$p=1/n=2\cdot 10^{-3}$.  Also included in the table are
our estimated error rates for the genie-aided decoders $D_0, D_1,
\ldots, D_5$.

\begin{table}[t]
    \setlength{\tabcolsep}{4pt}
    \centering
    \caption{Predicted Post-FEC BER for BCH Codes With and Without Miscorrection ($n=500,p=2\cdot10^{-3}$)}
    \begin{tabular}{|c|c|c|c|c|c|}\hline
    & $t=1$ & $t=2$ & $t=3$ & $t=4$ & $t=5$\\\hline
    Actual        & 1.78e-3 & 6.75e-4 & 1.77e-4 & 3.79e-5 & 7.18e-6\\
    $D_0$ Estimate & 1.26e-3 & 5.27e-4 & 1.60e-4 & 3.75e-5& 7.17e-6\\
    $D_1$ Estimate & 1.63e-3 & 6.44e-4 & 1.74e-4 & 3.84e-5& 7.21e-6\\
    $D_2$ Estimate & 1.74e-3 & 6.73e-4 & 1.77e-4 & 3.86e-5&        \\
    $D_3$ Estimate & 1.77e-3 & 6.79e-4 & 1.77e-4 &        &        \\
    $D_4$ Estimate & 1.78e-3 & 6.79e-4 &         &        &        \\
    $D_5$ Estimate & 1.78e-3 &         &         &        &        \\\hline
    \end{tabular}
    \label{tab:bsc-misc}
\end{table}

We see that for $t=1$ BCH codes (i.e., Hamming codes), the post-FEC BER
agrees closely with the simulation when we consider the effects of
miscorrection up to words of error weight $t+4=5$ (i.e., under
decoding with $D_4$).
For $t=2$ and $t=3$ BCH codes, the
predicted BERs under $D_2$ already closely match the simulation.
When $t \geq 4$, 
predicted BERs under $D_0$ (the miscorrection-free decoder)
closely agree with the simulated value.

We conclude that, in the crossover probability regime
of interest, adjusting error-rate estimates to take
miscorrections into account is important only for $t \leq 3$.  For
$t=1$, $2$, and $3$, we will consider miscorrections only
for received words of error weight up to $5$ (i.e., genie-aided
decoder $D_4$
for $t=1$, $D_3$ for $t=2$, and $D_2$ for $t=3$).   We will assume that
error weights greater than 5 lead to a decoding failure.

\begin{remark}
In the case of extended Hamming codes, miscorrections occur only when
the received error weight is odd. Using a similar argument as above, in
the event of a miscorrection, a word of odd weight $w > 2$ is assumed to
be miscorrected to a codeword of weight $w+1$.
\end{remark}

\begin{figure}[htbp]
    \centering
    \includegraphics{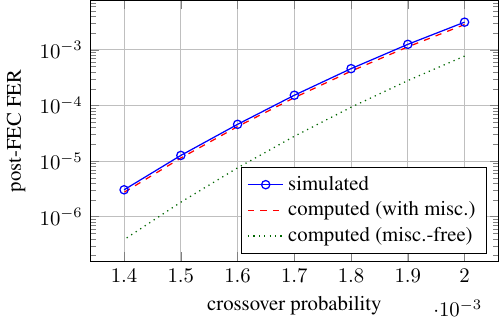}
\caption{FER versus crossover probability for 8 $\times$ RS(544, 514,
15) outer code and 64 $\times$ BCH(700, 680, 2) inner code with and
without miscorrection taken into account.  We take miscorrection into
account only for received BCH words of error weights 3, 4, and 5.}
    \label{fig:miscfree}
\end{figure}

Fig.~\ref{fig:miscfree} shows post-FEC FER curves comparing Monte Carlo
simulation with predicted FER using our analysis
\eqref{eq:bit-error-prob}--\eqref{eq:unionbound} with and without
miscorrections taken into account. The plot for the predicted FER with
miscorrection closely agrees with the plot obtained from the simulation.

Note that since we are interested in the post-FEC FER, not the BER, we
do not need to take the miscorrection rate of the outer RS codes into
account.  Any RS word receiving more than $T$ byte errors is already
sufficient to trigger a frame error.

\section{Performance-Complexity-Latency Trade-offs}
\label{sec:perfcomplattradeoffs}

\subsection{Metrics}
In this subsection, we will define and describe appropriate metrics
for performance, complexity, and latency of concatenated RS-BCH codes.

\subsubsection{Performance}
We define the \emph{performance} of a code as the gap to the Shannon
limit (in dB) for a binary symmetric channel at $10^{-13}$ post-FEC FER.
The binary symmetric channel is assumed to arise as
additive white Gaussian noise channel with antipodal
input under hard-decision detection, so a gap in dB measures
a noise-power ratio.
Let $R$ and $p^*$ denote the code rate and the channel crossover probability such
that the post-FEC FER is $10^{-13}$, respectively. The gap to the Shannon limit is then obtained as
\begin{align*}
\text{Gap (dB)}=20\log_{10}\left(\frac{\erfc^{-1}(2p^*)}{\erfc^{-1}\left(2\mathcal{H}^{-1}(1-R)\right)}\right),
\end{align*}
where $\erfc^{-1}$
denotes the inverse complementary error function,
and $\mathcal{H}^{-1}$ denotes the
inverse binary entropy function returning values in $[0,\frac{1}{2}]$.

Of course, other performance metrics, such as 
net coding gain (used in \cite{barakatain-lowldpcstc}) might be
considered.

\subsubsection{Complexity}
\label{sec:codecomplexity}

We define the \emph{complexity} to be the total number of elementary operations
performed by the inner and outer decoders in the worst case, normalized
by the number of information bits of the overall concatenated code.
As shown in Appendix~\ref{sec:elementaryops},
using a discrete logarithm-domain representation of finite field elements,
all decoding operations are implemented in terms of integer addition and subtraction
and table lookup.

There are five main RS/BCH decoding procedures to
consider:
\begin{enumerate}
\item syndrome computation (SC),
\item key equation solver (KE),
\item polynomial root-finding (RF),
\item error evaluation (EE) (for RS decoder only), and
\item bit correction (BC).
\end{enumerate}

A detailed analysis of the complexity of each decoding step is given in
Appendices~\ref{sec:decodercomplexity}--\ref{sec:rootfinding}
and the results are summarized in
Table~\ref{tab:bitops}.
For values of $t$ and $T$ smaller than 5, simplified procedures
for the KE and RF steps are assumed, as explained in
Appendices~\ref{sec:simplified}~and ~\ref{sec:rootfinding}. For values of $t$ and $T$ greater than 4,
general-purpose algorithms (Berlekamp-Massey and Chien search) are
assumed.

The complexity of an RS decoder is the sum of the complexities of the
SC, KE, RF, EE, and BC procedures, denoted as $\kappa_{\text{RS}}$.  The
complexity of a BCH decoder is the sum of the complexities of the SC,
KE, RF, and BC procedures, denoted as $\kappa_{\text{BCH}}$. Thus,
since the concatenated code
has $M$ RS decoders and $m$ BCH decoders with $MKB$ information bits,
the number of operations per decoded information bit is given as
\[
    \dfrac{M\kappa_{\text{RS}}+m\kappa_{\text{BCH}}}{MKB}.
\]

\begin{table}[t]
    \setlength{\tabcolsep}{2pt}
    \centering
    \caption{Number of Operations for RS and BCH Decoding, Worst Case}
    \begin{tabular}{|c|c|c|}\hline
        Operation & \# Ops. (RS) & \# Ops. (BCH) \\\hline
        SC & $6K(N-K)$ & $nt$ \\\hline
        KE ($T=1$) & $9$ & $0$ \\
        KE ($T=2$) & $54$ & $11$ \\
        KE ($T=3$) & $159$ & $23$ \\
        KE ($T=4$) & $336$ & $64$ \\
        KE ($T\geq 5$) & $2T(24T+8)$ & $2t(24t+8)$ \\\hline
        RF ($T=1$) & $0$ & $0$ \\
        RF ($T=2$) & $10$ & $10$ \\
        RF ($T=3$) & $37$ & $37$ \\
        RF ($T=4$) & $98$ & $98$ \\
        RF ($T\geq 5$) & $6NT$ & $6nt$ \\\hline
        EE & $T\left(6(2T+\lceil T/2 \rceil)-1\right)$ & ---\\\hline
        BC & $2T$ & $t$ \\\hline
    \end{tabular}
    \label{tab:bitops}
\end{table}

Note that making
different architectural or algorithmic assumptions may give
different formulas for the complexity measure.

\subsubsection{Latency}

Latency is typically defined as the length of time between when a
particular information bit arrives at the input of the encoder and when
its estimate leaves the output of the decoder.  When high-rate
systematic codes are used, bits are transmitted essentially as soon as
they arrive.  They undergo modulation, propagation over the channel,
channel equalization and receiver filtering,  etc., and then are
accumulated in a receiver buffer.  Decoding commences once the receiver
buffer is full.  Assuming a single decoder, the decoding time is
normally set equal to the time it takes to fill the receiver buffer,
since decoding more slowly than this is not possible (otherwise
codewords would accumulate faster than they can be decoded) and decoding
more quickly than this is not necessary (otherwise the decoder would
undergo idle periods).  Since signal-processing delays are usually
negligible compared with propagation and buffer-fill delays, in most
applications, the latency is dominated by the sum of the propagation
delay and double the time it takes to fill the receiver buffer.  Since
latter is proportional to the block length of the concatenated code, we
take the block length $mn$ as our latency metric.

\subsection{Code Search}
\label{sec:codesearch}

We are interested in determining combinations of
inner and outer code parameters leading to the best possible
tradeoffs between performance, complexity, and latency.
We first fix the target rate and an upper bound on latency
of our system.
We then sweep through the following
outer RS and inner BCH code parameters:
\begin{itemize}
\item outer code parameters: $5\leq B\leq 10$, $N\leq 2^B-1$, $0\leq T\leq 20$, $K=N-2T$;
\item inner code parameters: $7\leq b\leq 14$, $n\leq 2^{b}-1$,
$0\leq t\leq 15$, $k=n-bt$, $B\mid k$.  When $t=1$, we consider
using both Hamming codes and extended Hamming codes, and in
the latter case, $t=1\text{x}$ and $k=n-b-1$.  The case
$t=0$ corresponds to a system with $M$ outer RS codes 
and a trivial (rate one) inner code. Similarly, the case
$T=0$ corresponds to a system with $m$ inner BCH codes 
and a trivial (rate one) outer code.
\end{itemize}
We adjust the number of BCH codewords $m$ such that the latency
$mn$ is not greater than the latency bound.
We then compute $M=\left\lfloor\frac{mk}{NB}\right\rfloor$. In the
case of $MNB<mk$, we assume that the information positions of the
first $m-1$ BCH codewords are completely filled in with the bits from
the RS codewords. The remaining $mk-MNB$ information bits in the last
BCH codeword are zero-padded.
We also make sure that the code rate $R=MKB/mn$ is approximately the
same as the target rate. In our code search, we have $|R-\text{target
rate}|\leq 0.05$, i.e, we achieve a rate within $5\%$ of the
target rate.

Finally, the interleaving scheme is chosen such that in the adjacency
matrix $\mathbf{L}$, we have $L_{ij}\approx N/m$ (or $\approx k/BM$) for
all $i \in [M]^*$, $j \in [m]^*$.  This is achieved by picking the
entries of $\mathbf{L}$ in the set $\left\{\left\lfloor
N/m\right\rfloor, \left\lceil N/m\right\rceil\right\}$ such that
$\mathbf{L}$ satisfies \eqref{eq:sumofLs}.  This ensures that each RS
codeword contributes an (almost) equal number of bits to each BCH
codeword.

We note that in the case of a sufficiently high BCH decoding radius $t$
and for a sufficiently low code rate, the number of parity bits in a BCH
codeword of length $2^b-1$ can actually be fewer than $bt$ (or
equivalently, the code dimension can be higher than $n-bt$)
\cite{clark}. In this code search, we search for codes with rates
$0.85$ and $239/255\approx 0.937$,
and for BCH codes with such rates (or higher), we have verified
that $k=n-bt$ is indeed the highest possible BCH code dimension for a
given $n$, $b$, and $t$.

We emphasize that no Monte Carlo simulations are needed to evaluate the
various codes being considered.  Given the code parameters and channel
crossover probability, the FER is estimated using
\eqref{eq:bit-error-prob}--\eqref{eq:unionbound}.  We then determine the
appropriate crossover probability such that the post-FEC FER is
$10^{-13}$ and compute the gap to the Shannon limit.

\begin{figure}[t]
    \centering
    \includegraphics{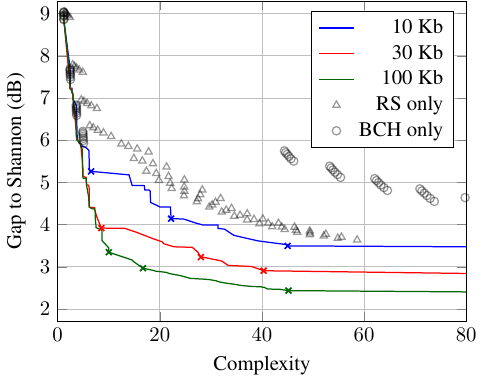}
\caption{Performance-complexity-latency trade-offs for target rate $\approx 0.85$.
Each curve shows Pareto-efficient operating
points corresponding to a fixed upper bound on
code latency as indicated.
For a latency bound of 10Kb, approximately 10,000 codes
were evaluated, among which 53 are Pareto-efficient.
For a latency bound of 30Kb, approximately 450,000 codes
were evaluated, among which 77 are Pareto-efficient.
For a latency bound of 100Kb, approximately $4 \times 10^6$ codes
were evaluated, among which 162 are Pareto-efficient.
The performance and complexity of RS codes and
BCH codes of rate $0.85$ and latency at most 10Kb are also included.
The points marked with ``$\times$'' correspond to the Pareto-efficient code parameters
listed in Table~\ref{tab:pareto85}.
}
    \label{fig:perfcomp510-85}
\end{figure}

\begin{figure}[t]
    \centering
    \includegraphics{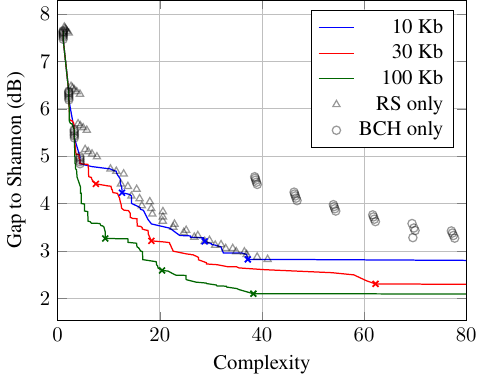}
\caption{Performance-complexity-latency trade-offs for target rate $239/255\approx 0.937$.
Each curve shows Pareto-efficient operating
points corresponding to a fixed upper bound on
code latency as indicated.
For a latency bound of 10Kb, approximately 8,000 codes
were evaluated, among which 64 are Pareto-efficient.
For a latency bound of 30Kb, approximately 370,000 codes
were evaluated, among which 109 are Pareto-efficient.
For a latency bound of 100Kb, approximately $5 \times 10^6$ codes
were evaluated, among which 114 are Pareto-efficient.
The performance and complexity of RS codes and
BCH codes of rate $0.937$ and latency at most 10Kb are also included.
The points marked with ``$\times$'' correspond to the Pareto-efficient code parameters
listed in Table~\ref{tab:pareto937}.
}
    \label{fig:perfcomp510}
\end{figure}

The Pareto optimal code parameters from the code search parameters
described above with rate targets of $0.85$ and $239/255 \approx 0.937$ are
shown in Figs.~\ref{fig:perfcomp510-85}~and~\ref{fig:perfcomp510}.
The parameters of a subset of the codes on the Pareto frontier are shown
in Tables~\ref{tab:pareto85}~and~\ref{tab:pareto937};
see Appendix~\ref{app:codeparameters} for a complete listing.

It is also interesting to note that for most of the code parameters on
the Pareto frontier, the
codes comprise weak ($t \leq 4$) inner
codes combined with  much stronger RS outer codes. At very low
complexities, we find that many $T=0$ codes (rate-1 outer codes, i.e., 
error protection only from the BCH codes) are Pareto efficient.

\begin{table}[t]
    \setlength{\tabcolsep}{2pt}
    \centering
    \caption{Some Pareto-Efficient Code Parameters, Rate $\approx 0.85$}
    \begin{tabular}{|c|c|c|c|c|c|c|c|c|c|c|}\hline
        $M$ & $N$ & $B$ & $T$ & $m$ & $n$ & $b$ & $t$ & Latency & Complexity & Gap (dB)\\\hline
        19 & 56 & 8 & 3 & 56 & 161 & 8 & 1x & 9016 & 6.6 & 5.26 \\
        3 & 274 & 10 & 10 & 92 & 98 & 7 & 1x & 9016 & 22.2 & 4.14 \\
        2 & 427 & 10 & 20 & 61 & 149 & 8 & 1x & 9089 & 45.0 & 3.50 \\
        27 & 120 & 8 & 3 & 130 & 224 & 8 & 3 & 29120 & 8.6 & 3.92 \\
        6 & 420 & 10 & 12 & 115 & 244 & 8 & 3 & 28060 & 28.0 & 3.23 \\
        3 & 860 & 10 & 18 & 89 & 326 & 9 & 4 & 29014 & 40.3 & 2.91 \\
        27 & 314 & 10 & 4 & 340 & 286 & 9 & 4 & 97240 & 10.0 & 3.35 \\
        29 & 300 & 10 & 6 & 300 & 326 & 9 & 4 & 97800 & 16.7 & 2.97 \\
        11 & 747 & 10 & 20 & 249 & 366 & 9 & 4 & 91134 & 45.2 & 2.43 \\
        \hline
    \end{tabular}
    \label{tab:pareto85}
\end{table}

\begin{table}[t]
    \setlength{\tabcolsep}{2pt}
    \centering
    \caption{Some Pareto-Efficient Code Parameters, Rate $\approx 0.937$}
    \begin{tabular}{|c|c|c|c|c|c|c|c|c|c|c|}\hline
        $M$ & $N$ & $B$ & $T$ & $m$ & $n$ & $b$ & $t$ & Latency & Complexity & Gap (dB)\\\hline
        2 & 391 & 10 & 6 & 17 & 472 & 11 & 1x & 8024 & 12.6 & 4.23 \\
        1 & 445 & 10 & 14 & --- & --- & --- & --- & 4450 & 28.8 & 3.20 \\
        1 & 692 & 10 & 18 & 1 & 6933 & 13 & 1 & 6933 & 37.2 & 2.82 \\
        16 & 215 & 8 & 4 & 35 & 806 & 13 & 1x & 28210 & 7.5 & 4.42 \\
        4 & 606 & 10 & 9 & 61 & 413 & 12 & 1x & 25193 & 18.3 & 3.21 \\
        3 & 960 & 10 & 20 & 67 & 440 & 10 & 1x & 29480 & 62.2 & 2.31 \\
        20 & 467 & 10 & 4 & 117 & 840 & 10 & 4 & 98280 & 9.3 & 3.26 \\
        13 & 646 & 10 & 9 & 86 & 1016 & 12 & 3 & 87376 & 20.4 & 2.59 \\
        9 & 952 & 10 & 18 & 68 & 1296 & 12 & 3 & 88128 & 38.3 & 2.10 \\
        \hline
    \end{tabular}
    \label{tab:pareto937}
\end{table}

\section{Conclusion}
\label{sec:conclusion}

In this paper, we have provided a formula to predict the frame error
rate of the concatenated RS-BCH codes. We have determined the parameters
of codes that are Pareto optimal with respect to particular measures of
performance, complexity, and latency.
The complexity measure simply counts the elementary operations (integer addition
and subtraction and table lookups) required for decoding.
Future work may include extending
the complexity measure to reflect hardware implementation or energy consumption.
Furthermore, it would be interesting to extend
the FER analysis to a pseudo-product-code-like structure, where there is
iterative decoding between the inner and outer codes.

\section{Acknowledgment}
The authors would like to thank the anonymous reviewers for their valuable comments
and suggestions.

\appendix
\subsection{Predicting the Post-FEC Bit Error Rate (BER)}
\label{sec:predictingser}

Suppose that the BCH and RS decoders are miscorrection-free.  Recall
from Sec.~\ref{sec:predictingfer} that $Y_i$ denotes the number of byte
($B$-bit symbol) errors at the input of the $i$th RS decoder.  Let
$\overline{Y}_{i}$ denote the number of byte errors at the output of the
$i$th RS decoder. Then,
\[
    \Pr(\overline{Y}_i=0)=\sum_{y=0}^T \Pr(Y_i=y)
\]
and
\begin{align*}
    \Pr(\overline{Y}_i=y)=\Pr(Y_i=y)~\text{for all}~y=T+1,\ldots,N.
\end{align*}
Thus, at the output of the $i$th RS decoder, we expect to see
\begin{align}
    E(\overline{Y}_i)=\sum_{y=T+1}^N yP(\overline{Y}_i=y)
    \label{eq:symberr-expected}
\end{align}
byte errors.

\begin{figure}[t]
    \centering
    \includegraphics{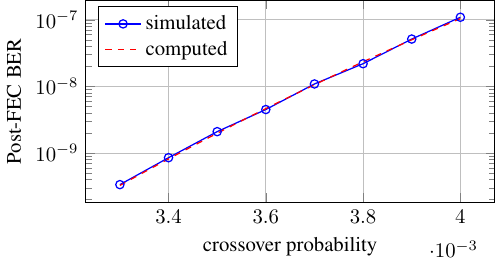}
\caption{BER versus crossover probability for $8\times~\text{RS}(544,514,15)$ outer code
and $68\times~\text{BCH}(690,640,5)$ BCH code. We assume that both outer
and inner decoders are miscorrection-free.
}
    \label{fig:berplot}
\end{figure}

We will now estimate the number of bit errors in the event of a byte
error. Let $Z$ denote the number of bit errors in a byte. For a binary
symmetric channel with crossover probability $p$, we have
\begin{align}
    \Pr(Z=z)=\binom{B}{z}p^z(1-p)^{B-z}\quad\text{for all}~z\in[B].
\end{align}
Thus, in the event of a byte error, we expect to see
\begin{align}
    E(Z\mid Z>0)=\sum_{z=1}^B z\frac{P(Z=z)}{P(Z>0)}
\end{align}
bit errors. Hence, the post-FEC BER is approximately
\begin{align}
    \frac{(E(\overline{Y}_1)+\cdots+E(\overline{Y}_M))E(Z\mid Z>0)}{MNB}.
    \label{eq:expectedBER}
\end{align}
Fig.~\ref{fig:berplot} shows post-FEC BER curves comparing Monte Carlo
simulation with predicted BER using our analysis \eqref{eq:symberr-expected}--\eqref{eq:expectedBER}.
Similar to those in Figs.~\ref{fig:misc}~and~\ref{fig:miscfree}, we observe
that the computed BERs agree closely with those computed \eqref{eq:expectedBER}.

\subsection{Finite Field Arithmetic}
\label{sec:elementaryops}

In this appendix we show that, by using a discrete logarithm representation,
finite field operations can be implemented using
integer addition and subtraction and table lookups, and we
derive the number of elementary operations needed.

Let $q=2^b$, and let $\alpha$ be a primitive element of $\mathbb{F}_q$.
Every nonzero element $x\in\mathbb{F}_q$ is of the form $x=\alpha^i$,
where $i\in[q-2]$. The exponent $i$ is the discrete logarithm of $x$
to the base $\alpha$, and we write $\log(x)=i$. For reasons that
will become clear, we define $\log(0)=2q-3$.

\subsubsection{Multiplication}
Let $x,y\in\mathbb{F}_{q}$ with $\log(x)=i$ and $\log(y)=j$ for
$i,j\in[q-2]\cup\{2q-3\}$. Then
$$\log(xy) = M(i+j),$$
where
$$
M(\ell)=\begin{cases}
\ell & \text{if}~0\leq \ell\leq q-2,\\
\ell-q+1 & \text{if}~q-1\leq \ell\leq 2q-4,\\
2q-3 & \text{if}~2q-3\leq\ell\leq 4q-6.
\end{cases}
$$
We assume that the values of $M(\ell)$ are stored in a table with $4q-5$ entries.
Thus, multiplying two elements in $\mathbb{F}_q$ comprises two elementary operations:
one integer addition and one table lookup. We note that multiplication by zero
could be handled conditionally; however, this table lookup approach avoids testing
whether $x$ or $y$ is zero, resulting in fewer operations. Since $(q-2)+(q-2)=2q-4$,
setting $\log(0)=2q-3$ (outside the range of the sum of logarithms
of nonzero elements)
minimizes the size of the $M$ table.

\subsubsection{Addition}
Let $x,y\in\mathbb{F}_{q}$ with $\log(x)=i$ and $\log(y)=j$ for
$i,j\in[q-2]\cup\{2q-3\}$. If $i,j\in[q-2]$, then
\begin{align}
    \log(x+y) &= \log\left(\alpha^i + \alpha^j\right) \nonumber\\
    &= \log\left(\alpha^i\left(1+\alpha^{j-i}\right)\right) \nonumber\\
    &= M\left(i+\log\left(1+\alpha^{j-i}\right)\right)\nonumber\\
    &= M(i+Z(j-i)), \label{eq:additiontable}
\end{align}
where
\begin{align*}
    Z(\ell)=\begin{cases}
        \ell & \text{if}~-2q+3\leq \ell\leq -q+1,\\
        \log(1+\alpha^\ell) & \text{if}~ -q+2\leq\ell\leq q-2,\\
        0 & \text{if}~q-1\leq \ell\leq 2q-3.
    \end{cases}
\end{align*}
We assume that the values of $Z(\ell)$, the Zech logarithm \cite{zech}
extended to include $\log(0)$,
are stored in a table with $4q-5$ entries,
designed so that \eqref{eq:additiontable} holds even when one or both
of $i$ and $j$ is $2q-3$.
As with multiplication, this avoids testing whether one
or both of the operands is the zero element.
Thus, adding two elements in $\mathbb{F}_q$ comprises four elementary
operations: one integer addition, one integer subtraction, and two table lookups.

\subsubsection{Division}
Let $x,y\in\mathbb{F}_{q}$ with $\log(x)=i$ and $\log(y)=j$ for
$i\in[q-2]\cup\{2q-3\}$ and $j\in[q-2]$. Then,
\begin{align*}
\log(x/y) &= M\left(\log(x)+\log(y^{-1})\right)\\
 &= M\left(i+I(j)\right),
\end{align*}
where $I(j) = -j\bmod(q-1)$. We assume that the values of $I(j)$ are stored
in a table with $q-1$ entries. The table lookup avoids performing a
$\bmod(q-1)$ operation. Thus, dividing $x$ by (a nonzero) $y$ comprises three
elementary operations: one integer addition and two table lookups.

\subsubsection{Converting Between Polynomial and Discrete Logarithm Representations}
Let $\alpha$ be a primitive element of $\mathbb{F}_{2^m}$. A \emph{polynomial representation}
of an element of $x\in\mathbb{F}_{2^m}$ is an expression of the form
$x=x_0+x_1\alpha+\cdots+x_{m-1}\alpha^{m-1}$, where usually the polynomial coefficients
are gathered into the $m$-tuple $(x_0,\ldots,x_{m-1})\in\mathbb{F}_2^m$. 
We assume that there exist lookup tables that can be used to convert between polynomial
and discrete logarithm representations of finite field elements.

\subsection{RS/BCH Decoder Complexity}
\label{sec:decodercomplexity}

\subsubsection{Syndrome Computation (SC)}

We suppose that the parity-check matrix for the RS code
is given in systematic form $H = [ -P^T \mid I ]$,
where $P$ is a $K \times (N-K)$ matrix and $I$ is an $(N-K) \times (N-K)$
identity matrix.  Computation of the syndrome
corresponding to an $N$-byte vector
then requires
$(N-K)$ inner-product operations of length $K$ vectors,
which requires $K(N-K)$ multiplications and $(K-1)(N-K)$
additions,
and additional
length $(N-K)$ vector addition, for a total
of $K(N-K)$ multiplications, and $K(N-K)$ additions.
This brings the number of elementary operations for SC to
$4K(N-K) + 2K(N-K)=6K(N-K)$.

The SC complexity for BCH codes, on the other hand, is composed of the XORs
of columns of the parity-check matrix according to the zero/one pattern in the
received word. 
We assume that each parity-check matrix
column comprises $t$ $b$-bit symbols over $\mathbb{F}_{2^b}$
in their polynomial
representation. Assuming that a symbol-wise XOR takes one operation,
at most $(n-1)t$ operations are needed in total to compute
the polynomial representation of the syndrome.
We then convert from polynomial representation to 
the discrete logarithm representation,
which takes another $t$ operations. In total, therefore,
the number of elementary operations required
for BCH syndrome computation is $nt$.

\subsubsection{Key Equation Solver (KE)}

There are two cases to consider to obtain the error-locator polynomial
and, in the case of RS decoding, error-evaluator polynomial.

\underline{\textbf{Case $T$ (or $t$) $= 1,2,3,4$:}} If the error correcting
radius of the RS decoder is small, the process of obtaining the coefficients of
the error-locator polynomial can be simplified by using Peterson's algorithm.
Suppose that the RS syndromes obtained from SC are $(S_1,S_2,\ldots,S_{2T})$.
The coefficients of the error-locator polynomial
\begin{align*}
    \Lambda(x)=1+\Lambda_1x+\cdots+\Lambda_\nu x^\nu
\end{align*}
for some $\nu\in [T]$ is obtained by solving a system of linear equations
\begin{align*}
    S_j=\sum_{i=1}^\nu \Lambda_iS_{j-i}\quad\text{for all}~j\in[\nu]^*.
\end{align*}
The worst-case complexity of the algorithm is attained when $\Lambda(x)$ has degree $T$, i.e., $\nu=T$.
We use Gaussian elimination to solve for $\Lambda_1,\ldots,\Lambda_T$.
The number of operations needed is:
\begin{itemize}
    \item $T=1$: 1 division,
    \item $T=2$: 3 additions, 3 multiplications, and 2 divisions,
    \item $T=3$: 12 additions, 15 multiplications, and 3 divisions,
    \item $T=4$: 30 additions, 36 multiplications, and 4 divisions.
\end{itemize}

In order to obtain the error-evaluator polynomial $\Omega(x)$ for RS decoders for $T=1,2,3,4$, we have
\begin{align*}
    \Omega(x)=&\Omega_0+\Omega_1x+\cdots+\Omega_{2T-1}x^{2T-1}\\
    \equiv&\Lambda(x)S(x)\bmod x^{2T},
\end{align*}
where $S(x)=S_1+S_2x+\cdots+S_{2t}x^{2t-1}$. The modulo operation above denotes
that terms of degree $2T$ and higher are discarded. Thus, assuming the
worst-case complexity that $\deg \Lambda(x)=T$, we have $\Omega_0=S_0$ and
\begin{align*}
\Omega_j=S_{j}+\sum_{i=1}^{\min\{j,T\}} \Lambda_i S_{j-i}~\quad\text{for all}~ j\in[2T-1]^*,
\end{align*}
which takes at most
$\frac{1}{2}T(3T-1)$ additions and $\frac{1}{2}T(3T-1)$ multiplications.
Thus, the complexity for computing the coefficients of $\Omega(x)$ is:
\begin{itemize}
    \item $T=1$: 1 addition and 1 multiplication,
    \item $T=2$: 5 additions and 5 multiplications,
    \item $T=3$: 12 additions and 12 multiplications,
    \item $T=4$: 22 additions and 22 multiplications.
\end{itemize}

The total complexity of KE for an RS decoder with $T=1,2,3,4$ is therefore:
\begin{itemize}
    \item $T=1$: $4 + 2 + 3 = 9$,
    \item $T=2$: $8\cdot 4 + 8\cdot 2 + 2\cdot 3 = 54$,
    \item $T=3$: $24\cdot 4 + 27\cdot 2 + 3\cdot 3 = 159$,
    \item $T=4$: $52\cdot 4 + 58\cdot 2 + 4\cdot 3 = 336$.
\end{itemize}

For BCH decoders, the algorithm can be simplified to even
fewer operations, as described in Appendix~\ref{sec:simplified}.

\underline{\textbf{Case $T$ (or $t$) $\geq 5$:}}
We consider the Berlekamp-Massey algorithm to obtain the error-locator and
error-evaluator polynomials.  We assume the reformulated inversionless
Berlekamp--Massey algorithm (RiBM) \cite{sarwate} with
$2T(3T+1)$ additions and $2T(6T+2)$ multiplications.
Using the same algorithm for BCH,
KE takes $2t(3t+1)$ additions and $2t(6t+2)$ multiplications.

\subsubsection{Polynomial Root-Finding (RF)}

There are two cases to consider to determine
the roots of the error-locator polynomial $\Lambda(x)$.

\underline{\textbf{Case $T$ (or $t$) $= 1,2,3,4$:}}
For small $T$ or $t$, the roots of $\Lambda(x)$ can be
determined by using lookup-table-based root-finding algorithms described in
Appendix~\ref{sec:rootfinding}.

\underline{\textbf{Case $T$ (or $t$) $\geq 5$:}}
We perform a brute-force Chien search of the roots of
$\Lambda(x)$.  In the worst case scenario, $\Lambda(x)$ has to be
evaluated $N$ times. Evaluating $\Lambda(x)$ with degree $T$, using Horner's
method, takes $T$ additions and $T$ multiplications, and so in total, the
worst-case complexity for RF is $NT(4+2)$. Similarly, the worst-case
complexity of RF for BCH is $nt(4+2)$.

\subsubsection{Error Evaluation (EE)}
\label{sec:errorevaluator}

The next part of RS decoding is to compute the error magnitude using Forney's
algorithm \cite{forney}.  The worst-case complexity of EE is when the
error-locator polynomial $\Lambda(x)$ has $T$ distinct roots, i.e.,
$\deg\Lambda(x)=T$.

Denote $\Lambda'(x)=\Lambda_1+2\Lambda_2x+\cdots+\nu\Lambda_{T-1}x^{T-1}$ to be
the formal derivative of $\Lambda(x)$.  Since the code operates over a field of
characteristic two, there are no odd-powered terms in $\Lambda'(x)$, leaving
$\Lambda'(x)$ with up to $\lceil T/2 \rceil$ nonzero even-powered terms.  Thus,
at every step of Horner's method of polynomial evaluation, one can use
$x^2$ instead of $x$ to reduce the number of additions and multiplications by
a factor of two.

Suppose that $X_{i_1}^{-1},\ldots,X_{i_T}^{-1}$ are the roots of $\Lambda(x)$
obtained in the RF step above.  Then, the error magnitude at position $i_j$
($j=1,\ldots,T$) is
\[
    E_j=-\frac{\Omega(X_{i_j}^{-1})}{\Lambda'(X_{i_j}^{-1})}.
\]
The complexity therefore comprises two polynomial evaluations via
Horner's method plus one finite field division.

Evaluating $\Omega(X_{i_j}^{-1})$ takes $2T-1$ finite field additions and
$2T-1$ multiplications. It takes one multiplication to compute
$(X_{i_j}^{-1})^2$, and evaluating $\Lambda'(X_{i_j}^{-1})$ takes $\lceil T/2
\rceil$ additions and $\lceil T/2 \rceil$ multiplications. Finally, we
divide $\Omega(X_{i_j}^{-1})$ by $\Lambda'(X_{i_j}^{-1})$. The total
complexity of EE is therefore $T\left((2T+\lceil T/2
\rceil - 1)(4+2) + 2 + 3\right)$.

\subsubsection{Bit Correction (BC)}

For a given RS word, there are up to $T$ bytes to be corrected. Assuming that there
are $T$ byte errors that are to be corrected, the $T$ error magnitudes are first converted
to the polynomial representation, which takes $T$ operations. Then, the bytes in the
$T$ locations of the received word prescribed by the RF are XOR-ed by the error magnitude.
Thus, it takes $2T$ operations.

On the other hand, there is no error evaluation needed for BCH decoding, and therefore the number
of operations for BC is at most $t$.

\subsection{Worst-Case Complexity of Simplified Peterson's Algorithm for $t=1,2,3,4$ BCH Decoders}
\label{sec:simplified}

This section describes the complexity of obtaining the coefficients of the
error locator polynomials using Peterson's algorithm for $t=2,3,4$ binary BCH
decoders with $n=2^b-1$ \cite{moon}. Throughout this section, assume that the
syndromes for $t=1$, $2$, $3$, and $4$ BCH words have form $(S_1)$,
$(S_1,S_3)$, $(S_1,S_3,S_5)$, and $(S_1,S_3,S_5,S_7)$, respectively, where
$S_1,S_3,S_5,S_7\in\mathbb{F}_{2^b}$. Also, for all positive integers $q$, let
$D_{q}=S_{q}+S_{1}^{q}$.
We also assume that division by zero never happens in the algorithm below.

\subsubsection{$t=1$}
The error-locator polynomial is
\[
\Lambda(x)=1+S_1 x.
\]
The coefficient of $x$ can be obtained directly from the syndromes. This
operation has zero complexity.

\subsubsection{$t=2$}
The error-locator polynomial is
\[
\Lambda(x)=1+S_1 x+\frac{D_3}{S_1}x^2.
\]
One addition and two multiplications are required to obtain
$D_3=S_1^3+S_3 = S_1\cdot S_1\cdot S_1+S_3$. In order to obtain
the coefficient of $x^2$, we perform one division.
In total, we have one addition,
two multiplications, and one inversion.
The complexity is therefore $4+2\cdot 2+3=11$

\subsubsection{$t=3$}
The error-locator polynomial is
\[
\Lambda(x) = 1 + \Lambda_1 x + \Lambda_2 x^2 + \Lambda_3 x^3,
\]
where
\[
\Lambda_1=S_1,~\Lambda_2=\frac{S_1^2S_3 + S_5}{D_3},~ \Lambda_3=D_3+S_1\Lambda_2.
\]
We first compute and store the values of $S_1^2$ and $S_1^3$,
each of which  requires
one multiplication. Then, computing $D_3$ takes one
addition. Using the value of $S_1^2$ and $D_3$ obtained in the previous step, computing $\Lambda_2$ takes one addition, one
multiplication, and one division.
Finally, computing $\Lambda_3$ takes one addition and one multiplication.
In total, we have
three additions, four multiplications, and one division.
The complexity is therefore $3\cdot 4 + 4\cdot 2 + 3 = 23$

\subsubsection{$t=4$}
The error-locator polynomial is
\[
\Lambda(x)=1+\Lambda_1x+\Lambda_2x^2+\Lambda_3x^3+\Lambda_4x^4,
\]
where
\begin{align*}
    \Lambda_1 = S_1,\quad 
    \Lambda_2 = \frac{S_1D_7 + S_3D_5}{S_3D_3+S_1D_5},
\end{align*}

\begin{align*}
    \Lambda_3 = D_3 + S_1\Lambda_2,\quad
    \Lambda_4 = \frac{S_5+S_1^2S_3+D_3\Lambda_2}{S_1}.
\end{align*}
We will compute and store a few intermediate terms listed in Table~\ref{tab:t4bch}.

\begin{table}[htbp]
    \setlength{\tabcolsep}{2pt}
    \centering
    \caption{Number of Operations of the Intermediate Terms for Computing the Coefficients of $\Lambda(x)$}
    \begin{tabular}{|c|c|c|}\hline
        Term & \# Additions & \# Multiplications\\\hline
        $S_1^2$ & $0$ & $1$\\\hline
        $S_1^4 = S_1^2\cdot S_1^2$ & $0$ & $1$\\\hline
        $S_1^6 = S_1^4\cdot S_1^2$ & $0$ & $1$\\\hline
        $D_3 = S_1^2\cdot S_1+S_3$ & $1$ & $1$\\\hline
        $D_5 = S_1^4\cdot S_1+S_5$ & $1$ & $1$\\\hline
        $D_7 = S_1^6\cdot S_1+S_5$ & $1$ & $1$\\\hline
    \end{tabular}
    \label{tab:t4bch}
\end{table}
The complexity to compute $\Lambda_2,\Lambda_3,\Lambda_4$ is listed below:
\begin{itemize}
    \item $\Lambda_2$: 2 additions, 4 multiplications, and 1 division,
    \item $\Lambda_3$: 1 addition and 1 multiplication,
    \item $\Lambda_4$: 2 additions, 2 multiplications, and 1 division.
\end{itemize}
In total, we have 8 additions, 13 multiplications, and 2 divisions. The complexity
is therefore $8\cdot 4 + 13\cdot 2 + 2\cdot 3 = 64$

\subsection{Worst-Case Complexity of Root-Finding Algorithms for Polynomials of Degree $\leq 4$ over $\mathbb{F}_{2^b}$}
\label{sec:rootfinding}
The root-finding algorithms for degree-1, 2, and 3 polynomials are based
on \cite{polkinghorn}. The root-finding algorithm for degree-4
polynomials is based on \cite{yan}. In these algorithms, aside from the lookup
tables specified in Appendix~\ref{sec:elementaryops}, we also store lookup
tables of:
\begin{itemize}
\item roots of $x^2+x+c=0$ for all $c\in\mathbb{F}_{2^b}$ (DP2),
\item roots of $x^3+x+c=0$ for all $c\in\mathbb{F}_{2^b}$ (DP3),
\item square roots (if they exist) for elements in $\mathbb{F}_{2^b}$ (SQRT).
\end{itemize}
to speed up calculations.
The worst case complexity for all cases is when the polynomials
have the same number of distinct roots as their degree.
We also assume that division by zero never happens in the algorithms below.

\subsubsection{Degree-1}
The root of $x+p=0$ is $x=p$. This operation is free and has zero complexity.

\subsubsection{Degree-2}
Using \cite{polkinghorn}, we determine the worst-case
complexity of computing the roots of
\begin{align}
x^2+px+q=0,
\label{eq:quadratics}
\end{align}
where $p,q\in\mathbb{F}_{2^b}$.  First, we compute $c=\frac{q}{p^2}$,
which takes 1 multiplication and 1 division. Then, we obtain the
roots $y_1$ and $y_2$ of $y^2+y+c=0$ using DP2. Thus, the roots of
\eqref{eq:quadratics} are $x_1=py_1$ and $x_2=py_2$. In total, we have 3
multiplications, 1 division, and 1 lookup, which bring the total worst-case
complexity to $3\cdot 2 + 3 + 1 = 10$.

\subsubsection{Degree-3}
Using \cite{polkinghorn}, we determine the worst-case
complexity of computing the roots of
\begin{align}
x^3+px^2+qx+r=0,
\label{eq:cubics}
\end{align}
where $p,q,r\in\mathbb{F}_{2^b}$. First, we compute $\eta=p^2+q$ and
$\delta=pq+r$, which take 2 additions and 2 multiplications. Then, we
compute
\begin{align}
c=\frac{\delta}{\eta^{3/2}}=\frac{\delta}{\eta\cdot\eta^{1/2}},
\label{eq:c3}
\end{align}
which
takes 1 multiplication, 1 division, and 1 lookup.
Next we compute the roots $y_1,y_2,y_3$ of $y^3+y+c=0$, using one DP3 lookup.
The roots of \eqref{eq:cubics} are therefore
$x_i=\eta^{1/2}y_i+p$ for all $i=1,2,3$.
Using the value of $\eta^{1/2}$ previously computed in \eqref{eq:c3}, this operation takes
3 additions and 3 multiplications.

In total, we have 5 additions, 6 multiplications, 1 division, and 2 lookups, which
bring the total worst-case complexity to $5\cdot 4 + 6\cdot 2 + 3 + 2 = 37$.

\subsubsection{Degree-4}
Using \cite{yan}, we determine the worst-case complexity
of computing the roots of
\begin{align}
x^4+px^3+qx^2+rx+s=0,
\label{eq:quartics}
\end{align}
where $p,q,r,s\in\mathbb{F}_{2^b}$. First, we compute
\begin{align}
a_0=s+\frac{qr}{p}+\left(\frac{r}{p}\right)^2\quad\text{and}\quad a_2=q+(pr)^{1/2},
\label{eq:a0anda2}
\end{align}
which requires 3 additions, 3 multiplications, 2 divisions, and 1 SQRT lookup.
Then, we compute
\[
k_2=\frac{a_0}{a_2^2}\quad\text{and}\quad k_0=p\left(\frac{k_2}{a_2}\right)^{1/2},
\]
which takes 2 multiplications, 2 divisions, and 1 SQRT lookup.
Next, we compute the roots $b_1,b_2,b_3$ of $b^3+b+k_1=0$ via 1 DP3 lookup.
Now, let
\begin{align}
\alpha=\frac{k_2}{1+b_1^4}=\frac{k_2}{1+(b_1\cdot b_1)^2},
\label{eq:alpha}
\end{align}
which takes 1 addition, 2
multiplications, and 1 division. Then, we take any one root $c_0$
of $c^2+c+\alpha=0$, which takes 1 DP2 lookup. Using the value of $b_1^2$ previously computed in
\eqref{eq:alpha}, we compute
$\beta=\left(1+\frac{1}{b_1^2}\right)c_0$, which takes 1 addition, 1
multiplication, and 1 inversion. Now, let $d_0$ be any root of
$d^2+d+\beta=0$, which takes 1 DP2 lookup.
Let $z_1=b_1d_0$, $z_2=z_1+b_1$, $z_3=z_1+b_2$, and
$z_4=z_1+z_2+z_3$. In total, we have 4 additions and 1 multiplication.
Finally,
\[
y_i=z_i\left(\frac{a_2}{a_0}\right)^{1/2}~\text{for all}~i=1,2,3,4,
\]
and the roots of \eqref{eq:quartics} are
\[
     x_i=\frac{1}{y_i}+\left(\frac{r}{p}\right)^{1/2} ~\text{for all}~i=1,2,3,4.
\]
Using the value of $\frac{r}{p}$ previously computed in \eqref{eq:a0anda2},
computing $y_i$ and $x_i$ ($i=1,2,3,4$) takes 4 additions, 4 multiplications, 1 division, and 6 lookups (2 for SQRT,
4 for inversion). The number of operations is summarized in
Table~\ref{tab:quartic}.

\begin{table}
\centering
\caption{Complexity of Root-Finding of A Quartic Polynomial}
\begin{tabular}{|c|c|c|c|c|}\hline
Term(s) & \# Add. & \# Mult. & \# Div. & \# Lookup\\\hline
$a_0$ & 2 & 2 & 2 & 0\\\hline
$a_2$ & 1 & 1 & 0 & 1\\\hline
$k_2$ & 0 & 1 & 1 & 0\\\hline
$k_0$ & 0 & 1 & 1 & 1\\\hline
$b_1,b_2,b_3$ & 0 & 0 & 0 & 1\\\hline
$\alpha$ & 1 & 2 & 1 & 0\\\hline
$c_0$ & 0 & 0 & 0 & 1\\\hline
$\beta$ & 1 & 1 & 0 & 1\\\hline
$d_0$ & 0 & 0 & 0 & 1\\\hline
$z_1,z_2,z_3,z_4$ & 4 & 1 & 0 & 0\\\hline
$y_1,y_2,y_3,y_4$ & 0 & 4 & 1 & 1\\\hline
$x_1,x_2,x_3,x_4$ & 4 & 0 & 0 & 5\\\hline
\end{tabular}
\label{tab:quartic}
\end{table}

Hence, the total number of operations is 13 additions, 13
multiplications, 6 divisions, and 12 lookups, which brings the worst-case complexity
to $13\cdot 4 + 13\cdot 2 + 6 \cdot 3 + 12 = 98$.

\subsection{Pareto-Efficient Code Parameters}
\label{app:codeparameters}
Tables~\ref{tab:10k-rate085}--\ref{tab:100k-rate0937} list the parameters of Pareto-efficient codes
shown in Figs.~\ref{fig:perfcomp510-85}~and~\ref{fig:perfcomp510}.
Here, ``$t=\text{1}$'' and ``$t=\text{1x}$'' denote inner Hamming codes and extended Hamming codes, respectively.
Codes with rate-1 outer codes have ``---'' for their $M$, $N$, $B$, and $T$. Similarly, codes with rate-1 inner
codes have ``---'' for their $m$, $n$, $b$, and $t$.

\begin{table*}
\setlength{\tabcolsep}{2pt}
\centering
\caption{Pareto-Efficient Code Parameters With Latency $\leq 10,000$ Bits and Rate $\approx 0.85$}
\begin{tabular}{|c|c|c|c|c|c|c|c|c|c|c||c|c|c|c|c|c|c|c|c|c|c|c|c|}
    \hline
        $M$ & $N$ & $B$ & $T$ & $m$ & $n$ & $b$ & $t$ & Latency & Complexity & Gap (dB) & $M$ & $N$ & $B$ & $T$ & $m$ & $n$ & $b$ & $t$ & Latency & Complexity & Gap (dB)\\\hline
        --- & --- & --- & --- & 1 & 94 & 14 & 1 & 94 & 1.2 & 9.04 & 23 & 59 & 6 & 2 & 124 & 73 & 7 & 1 & 9052 & 5.6 & 5.82 \\
        --- & --- & --- & --- & 1 & 87 & 13 & 1 & 87 & 1.2 & 9.03 & 12 & 75 & 7 & 2 & 43 & 163 & 8 & 2 & 7009 & 6.1 & 5.75 \\
        --- & --- & --- & --- & 1 & 74 & 11 & 1 & 74 & 1.2 & 8.97 & 9 & 88 & 8 & 3 & 88 & 79 & 7 & 1 & 6952 & 6.2 & 5.27 \\
        --- & --- & --- & --- & 1 & 67 & 10 & 1 & 67 & 1.2 & 8.96 & 19 & 56 & 8 & 3 & 56 & 161 & 8 & 1x & 9016 & 6.6 & 5.26 \\
        --- & --- & --- & --- & 1 & 54 & 8 & 1 & 54 & 1.2 & 8.88 & 6 & 171 & 8 & 5 & 103 & 88 & 7 & 1x & 9064 & 14.0 & 5.18 \\
        --- & --- & --- & --- & 1 & 47 & 7 & 1 & 47 & 1.2 & 8.86 & 7 & 120 & 10 & 6 & 60 & 149 & 8 & 1x & 8940 & 14.6 & 4.93 \\
        --- & --- & --- & --- & 1 & 187 & 14 & 2 & 187 & 2.4 & 7.70 & 5 & 134 & 10 & 7 & 24 & 293 & 12 & 1x & 7032 & 17.0 & 4.92 \\
        --- & --- & --- & --- & 1 & 174 & 13 & 2 & 174 & 2.4 & 7.67 & 5 & 160 & 8 & 6 & 89 & 79 & 7 & 1 & 7031 & 17.1 & 4.82 \\
        7 & 74 & 10 & 1 & 87 & 69 & 8 & 1x & 6003 & 2.4 & 7.23 & 9 & 120 & 8 & 6 & 47 & 193 & 8 & 1x & 9071 & 17.9 & 4.82 \\
        34 & 30 & 8 & 1 & 93 & 97 & 8 & 1x & 9021 & 2.8 & 7.23 & 1 & 694 & 10 & 7 & 24 & 334 & 11 & 4 & 8016 & 18.1 & 4.79 \\
        19 & 30 & 8 & 1 & 52 & 97 & 8 & 1x & 5044 & 2.8 & 7.17 & 4 & 207 & 10 & 8 & 83 & 109 & 8 & 1x & 9047 & 18.2 & 4.52 \\
        43 & 22 & 8 & 1 & 50 & 161 & 8 & 1x & 8050 & 2.9 & 7.12 & 7 & 166 & 8 & 7 & 83 & 120 & 7 & 1x & 9960 & 20.1 & 4.42 \\
        7 & 122 & 7 & 1 & 143 & 49 & 7 & 1 & 7007 & 3.0 & 7.11 & 4 & 187 & 10 & 8 & 11 & 728 & 12 & 4 & 8008 & 22.1 & 4.41 \\
        24 & 38 & 7 & 1 & 83 & 85 & 7 & 1x & 7055 & 3.0 & 7.02 & 3 & 274 & 10 & 10 & 92 & 98 & 7 & 1x & 9016 & 22.2 & 4.14 \\
        17 & 38 & 7 & 1 & 59 & 85 & 7 & 1x & 5015 & 3.0 & 6.97 & 2 & 407 & 10 & 11 & 37 & 244 & 12 & 2 & 9028 & 24.6 & 4.13 \\
        --- & --- & --- & --- & 1 & 280 & 14 & 3 & 280 & 3.6 & 6.83 & 3 & 280 & 10 & 12 & 56 & 161 & 10 & 1x & 9016 & 27.1 & 4.04 \\
        --- & --- & --- & --- & 1 & 260 & 13 & 3 & 260 & 3.6 & 6.80 & 1 & 620 & 10 & 12 & 20 & 350 & 10 & 4 & 7000 & 28.4 & 3.99 \\
        --- & --- & --- & --- & 1 & 240 & 12 & 3 & 240 & 3.7 & 6.76 & 5 & 235 & 8 & 11 & 47 & 212 & 11 & 1x & 9964 & 31.4 & 3.99 \\
        --- & --- & --- & --- & 1 & 220 & 11 & 3 & 220 & 3.7 & 6.73 & 2 & 327 & 10 & 14 & 47 & 150 & 9 & 1x & 7050 & 31.4 & 3.91 \\
        --- & --- & --- & --- & 1 & 200 & 10 & 3 & 200 & 3.7 & 6.69 & 1 & 880 & 10 & 15 & 44 & 227 & 9 & 3 & 9988 & 32.7 & 3.89 \\
        --- & --- & --- & --- & 1 & 180 & 9 & 3 & 180 & 3.7 & 6.64 & 2 & 367 & 10 & 14 & 15 & 534 & 11 & 4 & 8010 & 34.5 & 3.75 \\
        --- & --- & --- & --- & 1 & 160 & 8 & 3 & 160 & 3.7 & 6.59 & 2 & 414 & 10 & 16 & 83 & 109 & 8 & 1x & 9047 & 35.1 & 3.73 \\
        12 & 67 & 10 & 2 & 67 & 133 & 12 & 1x & 8911 & 3.8 & 6.15 & 1 & 800 & 10 & 18 & 32 & 280 & 10 & 3 & 8960 & 39.2 & 3.61 \\
        14 & 47 & 10 & 2 & 47 & 150 & 9 & 1x & 7050 & 3.9 & 6.01 & 2 & 427 & 10 & 20 & 45 & 201 & 10 & 1x & 9045 & 45.0 & 3.52 \\
        19 & 66 & 7 & 1 & 66 & 151 & 9 & 2 & 9966 & 4.2 & 5.93 & 2 & 427 & 10 & 20 & 61 & 149 & 8 & 1x & 9089 & 45.0 & 3.50 \\
        21 & 56 & 6 & 1 & 59 & 136 & 8 & 2 & 8024 & 4.6 & 5.91 & 1 & 800 & 10 & 18 & 16 & 560 & 10 & 6 & 8960 & 88.7 & 3.47 \\
        11 & 71 & 7 & 2 & 71 & 85 & 7 & 1x & 6035 & 4.9 & 5.86 & & & & & & & & & & &  \\
        \hline
\end{tabular}

\vspace{0.1cm}
There are 10,055 data points with 53 of them on the Pareto frontier.
\label{tab:10k-rate085}
\end{table*}

\begin{table*}[htbp]
\setlength{\tabcolsep}{2pt}
\centering
\caption{Pareto-Efficient Code Parameters With Latency $\leq 30,000$ Bits and Rate $\approx 0.85$}
\begin{tabular}{|c|c|c|c|c|c|c|c|c|c|c||c|c|c|c|c|c|c|c|c|c|c|c|c|}
    \hline
        $M$ & $N$ & $B$ & $T$ & $m$ & $n$ & $b$ & $t$ & Latency & Complexity & Gap (dB) & $M$ & $N$ & $B$ & $T$ & $m$ & $n$ & $b$ & $t$ & Latency & Complexity & Gap (dB)\\\hline
        --- & --- & --- & --- & 1 & 94 & 14 & 1 & 94 & 1.2 & 9.04 & 31 & 54 & 10 & 1 & 84 & 227 & 9 & 3 & 19068 & 4.9 & 5.34 \\
        --- & --- & --- & --- & 1 & 87 & 13 & 1 & 87 & 1.2 & 9.03 & 19 & 107 & 10 & 2 & 113 & 204 & 12 & 2 & 23052 & 5.0 & 5.21 \\
        --- & --- & --- & --- & 1 & 74 & 11 & 1 & 74 & 1.2 & 8.97 & 19 & 107 & 10 & 2 & 120 & 192 & 11 & 2 & 23040 & 5.0 & 5.14 \\
        --- & --- & --- & --- & 1 & 67 & 10 & 1 & 67 & 1.2 & 8.96 & 23 & 139 & 8 & 2 & 160 & 182 & 11 & 2 & 29120 & 5.5 & 5.12 \\
        --- & --- & --- & --- & 1 & 54 & 8 & 1 & 54 & 1.2 & 8.88 & 30 & 107 & 8 & 2 & 169 & 172 & 10 & 2 & 29068 & 5.6 & 5.11 \\
        --- & --- & --- & --- & 1 & 47 & 7 & 1 & 47 & 1.2 & 8.86 & 23 & 115 & 8 & 2 & 156 & 154 & 9 & 2 & 24024 & 5.6 & 5.04 \\
        7 & 274 & 10 & 1 & 274 & 81 & 10 & 1x & 22194 & 2.4 & 7.34 & 27 & 99 & 8 & 2 & 191 & 126 & 7 & 2 & 24066 & 5.6 & 4.97 \\
        5 & 434 & 10 & 1 & 434 & 58 & 7 & 1x & 25172 & 2.4 & 7.32 & 19 & 99 & 8 & 2 & 135 & 126 & 7 & 2 & 17010 & 5.6 & 4.94 \\
        5 & 414 & 10 & 1 & 414 & 58 & 7 & 1x & 24012 & 2.4 & 7.31 & 23 & 99 & 7 & 2 & 143 & 126 & 7 & 2 & 18018 & 6.1 & 4.93 \\
        5 & 294 & 10 & 1 & 294 & 58 & 7 & 1x & 17052 & 2.4 & 7.28 & 16 & 95 & 7 & 2 & 95 & 126 & 7 & 2 & 11970 & 6.1 & 4.92 \\
        13 & 154 & 10 & 1 & 334 & 69 & 9 & 1 & 23046 & 2.4 & 7.22 & 27 & 94 & 10 & 1 & 94 & 310 & 10 & 4 & 29140 & 6.2 & 4.84 \\
        13 & 154 & 10 & 1 & 334 & 69 & 8 & 1x & 23046 & 2.4 & 7.25 & 19 & 107 & 10 & 2 & 113 & 204 & 8 & 3 & 23052 & 6.2 & 4.41 \\
        33 & 54 & 8 & 1 & 128 & 126 & 13 & 1x & 16128 & 2.8 & 7.22 & 31 & 83 & 8 & 2 & 103 & 224 & 8 & 3 & 23072 & 6.9 & 4.41 \\
        61 & 46 & 8 & 1 & 234 & 107 & 10 & 1x & 25038 & 2.8 & 7.21 & 19 & 152 & 8 & 3 & 181 & 144 & 8 & 2 & 26064 & 7.2 & 4.40 \\
        49 & 46 & 8 & 1 & 188 & 107 & 10 & 1x & 20116 & 2.8 & 7.19 & 19 & 152 & 8 & 3 & 207 & 126 & 7 & 2 & 26082 & 7.3 & 4.37 \\
        44 & 46 & 8 & 1 & 169 & 107 & 10 & 1x & 18083 & 2.8 & 7.15 & 28 & 95 & 7 & 2 & 95 & 220 & 8 & 3 & 20900 & 7.3 & 4.34 \\
        43 & 22 & 8 & 1 & 50 & 161 & 8 & 1x & 8050 & 2.9 & 7.12 & 15 & 154 & 10 & 4 & 154 & 168 & 9 & 2 & 25872 & 7.7 & 4.16 \\
        7 & 122 & 7 & 1 & 143 & 49 & 7 & 1 & 7007 & 3.0 & 7.11 & 27 & 120 & 8 & 3 & 130 & 224 & 8 & 3 & 29120 & 8.6 & 3.92 \\
        24 & 38 & 7 & 1 & 83 & 85 & 7 & 1x & 7055 & 3.0 & 7.02 & 18 & 127 & 10 & 5 & 127 & 196 & 8 & 2 & 24892 & 13.1 & 3.91 \\
        17 & 38 & 7 & 1 & 59 & 85 & 7 & 1x & 5015 & 3.0 & 6.97 & 20 & 147 & 8 & 5 & 147 & 176 & 8 & 2 & 25872 & 15.5 & 3.81 \\
        5 & 487 & 10 & 2 & 487 & 58 & 7 & 1x & 28246 & 3.6 & 6.25 & 13 & 232 & 8 & 6 & 121 & 224 & 8 & 3 & 27104 & 18.8 & 3.66 \\
        5 & 467 & 10 & 2 & 467 & 58 & 7 & 1x & 27086 & 3.6 & 6.24 & 7 & 360 & 10 & 9 & 180 & 156 & 8 & 2 & 28080 & 20.5 & 3.59 \\
        6 & 307 & 10 & 2 & 307 & 69 & 8 & 1x & 21183 & 3.6 & 6.15 & 7 & 367 & 10 & 8 & 89 & 326 & 9 & 4 & 29014 & 20.8 & 3.55 \\
        6 & 247 & 10 & 2 & 247 & 69 & 8 & 1x & 17043 & 3.6 & 6.14 & 6 & 400 & 10 & 9 & 83 & 326 & 9 & 4 & 27058 & 22.8 & 3.47 \\
        11 & 234 & 10 & 1 & 234 & 128 & 9 & 2 & 29952 & 3.7 & 6.13 & 5 & 500 & 10 & 12 & 167 & 168 & 9 & 2 & 28056 & 26.4 & 3.46 \\
        12 & 194 & 10 & 1 & 233 & 116 & 8 & 2 & 27028 & 3.7 & 6.06 & 5 & 520 & 10 & 12 & 104 & 280 & 10 & 3 & 29120 & 27.5 & 3.34 \\
        15 & 67 & 10 & 2 & 101 & 110 & 9 & 1x & 11110 & 3.8 & 6.02 & 6 & 420 & 10 & 12 & 115 & 244 & 8 & 3 & 28060 & 28.0 & 3.23 \\
        14 & 47 & 10 & 2 & 47 & 150 & 9 & 1x & 7050 & 3.9 & 6.01 & 4 & 600 & 10 & 12 & 83 & 326 & 9 & 4 & 27058 & 28.5 & 3.23 \\
        41 & 46 & 8 & 1 & 135 & 126 & 7 & 2 & 17010 & 4.1 & 5.99 & 5 & 487 & 10 & 14 & 111 & 244 & 8 & 3 & 27084 & 32.1 & 3.14 \\
        19 & 66 & 7 & 1 & 66 & 151 & 9 & 2 & 9966 & 4.2 & 5.93 & 5 & 487 & 10 & 14 & 74 & 366 & 9 & 4 & 27084 & 33.4 & 3.03 \\
        21 & 66 & 7 & 1 & 73 & 151 & 9 & 2 & 11023 & 4.2 & 5.93 & 4 & 627 & 10 & 17 & 105 & 267 & 9 & 3 & 28035 & 38.0 & 3.01 \\
        17 & 67 & 8 & 2 & 114 & 88 & 7 & 1x & 10032 & 4.4 & 5.92 & 3 & 860 & 10 & 18 & 89 & 326 & 9 & 4 & 29014 & 40.3 & 2.91 \\
        17 & 67 & 8 & 2 & 127 & 79 & 7 & 1 & 10033 & 4.4 & 5.92 & 5 & 534 & 10 & 19 & 47 & 620 & 10 & 5 & 29140 & 83.3 & 2.84 \\
        26 & 62 & 6 & 1 & 81 & 136 & 8 & 2 & 11016 & 4.5 & 5.90 & 3 & 747 & 10 & 20 & 59 & 425 & 9 & 5 & 25075 & 85.2 & 2.82 \\
        84 & 46 & 6 & 1 & 155 & 168 & 9 & 2 & 26040 & 4.5 & 5.87 & 4 & 594 & 10 & 19 & 33 & 790 & 10 & 7 & 26070 & 100.2 & 2.81 \\
        26 & 74 & 10 & 1 & 77 & 286 & 12 & 3 & 22022 & 4.9 & 5.44 & 4 & 654 & 10 & 19 & 41 & 710 & 10 & 7 & 29110 & 100.7 & 2.79 \\
        32 & 74 & 10 & 1 & 103 & 263 & 11 & 3 & 27089 & 4.9 & 5.41 & 3 & 612 & 10 & 18 & 38 & 540 & 10 & 5 & 20520 & 182.9 & 2.78 \\
        32 & 74 & 10 & 1 & 113 & 240 & 10 & 3 & 27120 & 4.9 & 5.37 & 5 & 544 & 10 & 20 & 48 & 620 & 10 & 5 & 29760 & 188.5 & 2.63 \\
        44 & 54 & 10 & 1 & 119 & 227 & 9 & 3 & 27013 & 4.9 & 5.36 & & & & & & & & & & &  \\
        \hline
\end{tabular}

\vspace{0.1cm}
There are 449,981 data points with 77 of them on the Pareto frontier.
\label{tab:30k-rate085}
\end{table*}

\begin{table*}[htbp]
\setlength{\tabcolsep}{2pt}
\renewcommand{\arraystretch}{0.88}
\centering
\caption{Pareto-Efficient Code Parameters With Latency $\leq 100,000$ Bits and Rate $\approx 0.85$}
\begin{tabular}{|c|c|c|c|c|c|c|c|c|c|c||c|c|c|c|c|c|c|c|c|c|c|}
    \hline
        $M$ & $N$ & $B$ & $T$ & $m$ & $n$ & $b$ & $t$ & Latency & Complexity & Gap (dB) & $M$ & $N$ & $B$ & $T$ & $m$ & $n$ & $b$ & $t$ & Latency & Complexity & Gap (dB)\\\hline
        --- & --- & --- & --- & 1 & 94 & 14 & 1 & 94 & 1.2 & 9.04 & 19 & 427 & 10 & 2 & 812 & 116 & 8 & 2 & 94192 & 4.9 & 5.20 \\
        --- & --- & --- & --- & 1 & 87 & 13 & 1 & 87 & 1.2 & 9.03 & 10 & 407 & 10 & 2 & 407 & 116 & 8 & 2 & 47212 & 4.9 & 5.15 \\
        --- & --- & --- & --- & 1 & 74 & 11 & 1 & 74 & 1.2 & 8.97 & 10 & 387 & 10 & 2 & 387 & 116 & 8 & 2 & 44892 & 4.9 & 5.15 \\
        --- & --- & --- & --- & 1 & 67 & 10 & 1 & 67 & 1.2 & 8.96 & 10 & 327 & 10 & 2 & 327 & 116 & 8 & 2 & 37932 & 4.9 & 5.14 \\
        --- & --- & --- & --- & 1 & 54 & 8 & 1 & 54 & 1.2 & 8.88 & 10 & 267 & 10 & 2 & 267 & 116 & 8 & 2 & 30972 & 4.9 & 5.13 \\
        --- & --- & --- & --- & 1 & 47 & 7 & 1 & 47 & 1.2 & 8.86 & 36 & 87 & 10 & 2 & 209 & 168 & 9 & 2 & 35112 & 5.0 & 5.10 \\
        9 & 934 & 10 & 1 & 1051 & 93 & 13 & 1 & 97743 & 2.4 & 7.60 & 48 & 115 & 8 & 2 & 325 & 154 & 9 & 2 & 50050 & 5.6 & 5.09 \\
        9 & 934 & 10 & 1 & 1051 & 93 & 12 & 1x & 97743 & 2.4 & 7.60 & 23 & 115 & 8 & 2 & 156 & 154 & 9 & 2 & 24024 & 5.6 & 5.04 \\
        9 & 894 & 10 & 1 & 1006 & 93 & 13 & 1 & 93558 & 2.4 & 7.58 & 36 & 99 & 8 & 2 & 223 & 144 & 8 & 2 & 32112 & 5.6 & 5.01 \\
        9 & 894 & 10 & 1 & 1006 & 93 & 12 & 1x & 93558 & 2.4 & 7.59 & 27 & 99 & 8 & 2 & 191 & 126 & 7 & 2 & 24066 & 5.6 & 4.97 \\
        10 & 794 & 10 & 1 & 993 & 93 & 13 & 1 & 92349 & 2.4 & 7.58 & 19 & 99 & 8 & 2 & 135 & 126 & 7 & 2 & 17010 & 5.6 & 4.94 \\
        10 & 794 & 10 & 1 & 993 & 93 & 12 & 1x & 92349 & 2.4 & 7.58 & 116 & 67 & 8 & 2 & 371 & 186 & 9 & 2 & 69006 & 5.7 & 4.91 \\
        11 & 714 & 10 & 1 & 982 & 93 & 13 & 1 & 91326 & 2.4 & 7.57 & 25 & 167 & 10 & 2 & 167 & 286 & 12 & 3 & 47762 & 6.1 & 4.62 \\
        11 & 714 & 10 & 1 & 982 & 93 & 12 & 1x & 91326 & 2.4 & 7.58 & 23 & 167 & 10 & 2 & 167 & 263 & 11 & 3 & 43921 & 6.1 & 4.57 \\
        11 & 734 & 10 & 1 & 1010 & 93 & 12 & 1x & 93930 & 2.4 & 7.56 & 21 & 167 & 10 & 2 & 167 & 240 & 10 & 3 & 40080 & 6.1 & 4.51 \\
        15 & 554 & 10 & 1 & 1039 & 93 & 12 & 1x & 96627 & 2.4 & 7.43 & 64 & 127 & 10 & 2 & 407 & 227 & 9 & 3 & 92389 & 6.2 & 4.50 \\
        7 & 274 & 10 & 1 & 274 & 81 & 10 & 1x & 22194 & 2.4 & 7.34 & 25 & 127 & 10 & 2 & 159 & 227 & 9 & 3 & 36093 & 6.2 & 4.45 \\
        5 & 434 & 10 & 1 & 434 & 58 & 7 & 1x & 25172 & 2.4 & 7.32 & 19 & 107 & 10 & 2 & 113 & 204 & 8 & 3 & 23052 & 6.2 & 4.41 \\
        5 & 414 & 10 & 1 & 414 & 58 & 7 & 1x & 24012 & 2.4 & 7.31 & 63 & 131 & 8 & 2 & 230 & 327 & 13 & 3 & 75210 & 6.7 & 4.38 \\
        5 & 294 & 10 & 1 & 294 & 58 & 7 & 1x & 17052 & 2.4 & 7.28 & 19 & 152 & 8 & 3 & 207 & 126 & 7 & 2 & 26082 & 7.3 & 4.37 \\
        13 & 154 & 10 & 1 & 334 & 69 & 9 & 1 & 23046 & 2.4 & 7.22 & 28 & 95 & 7 & 2 & 95 & 220 & 8 & 3 & 20900 & 7.3 & 4.34 \\
        13 & 154 & 10 & 1 & 334 & 69 & 8 & 1x & 23046 & 2.4 & 7.25 & 34 & 247 & 10 & 2 & 247 & 392 & 13 & 4 & 96824 & 7.3 & 4.30 \\
        17 & 198 & 8 & 1 & 421 & 74 & 10 & 1 & 31154 & 2.7 & 7.18 & 31 & 260 & 10 & 3 & 299 & 309 & 13 & 3 & 92391 & 7.4 & 4.21 \\
        129 & 46 & 8 & 1 & 457 & 116 & 12 & 1 & 53012 & 2.8 & 7.16 & 32 & 260 & 10 & 3 & 333 & 286 & 12 & 3 & 95238 & 7.4 & 4.17 \\
        44 & 46 & 8 & 1 & 169 & 107 & 10 & 1x & 18083 & 2.8 & 7.15 & 31 & 260 & 10 & 3 & 323 & 286 & 12 & 3 & 92378 & 7.4 & 4.17 \\
        231 & 22 & 8 & 1 & 268 & 161 & 9 & 1 & 43148 & 2.8 & 7.15 & 34 & 240 & 10 & 3 & 355 & 263 & 11 & 3 & 93365 & 7.4 & 4.13 \\
        43 & 22 & 8 & 1 & 50 & 161 & 8 & 1x & 8050 & 2.9 & 7.12 & 34 & 240 & 10 & 3 & 389 & 240 & 10 & 3 & 93360 & 7.4 & 4.07 \\
        7 & 122 & 7 & 1 & 143 & 49 & 7 & 1 & 7007 & 3.0 & 7.11 & 25 & 167 & 10 & 2 & 167 & 286 & 9 & 4 & 47762 & 7.4 & 4.04 \\
        24 & 38 & 7 & 1 & 83 & 85 & 7 & 1x & 7055 & 3.0 & 7.02 & 52 & 160 & 10 & 3 & 463 & 204 & 8 & 3 & 94452 & 7.5 & 3.98 \\
        17 & 38 & 7 & 1 & 59 & 85 & 7 & 1x & 5015 & 3.0 & 6.97 & 21 & 160 & 10 & 3 & 187 & 204 & 8 & 3 & 38148 & 7.5 & 3.93 \\
        8 & 1007 & 10 & 2 & 1007 & 93 & 13 & 1 & 93651 & 3.6 & 6.54 & 27 & 248 & 8 & 3 & 319 & 192 & 8 & 3 & 61248 & 8.4 & 3.89 \\
        8 & 1007 & 10 & 2 & 1007 & 93 & 12 & 1x & 93651 & 3.6 & 6.54 & 23 & 248 & 8 & 3 & 272 & 192 & 8 & 3 & 52224 & 8.4 & 3.88 \\
        8 & 987 & 10 & 2 & 987 & 93 & 12 & 1x & 91791 & 3.6 & 6.53 & 41 & 200 & 10 & 3 & 216 & 432 & 13 & 4 & 93312 & 8.7 & 3.87 \\
        9 & 887 & 10 & 2 & 998 & 93 & 12 & 1x & 92814 & 3.6 & 6.52 & 26 & 314 & 10 & 4 & 355 & 263 & 11 & 3 & 93365 & 8.7 & 3.81 \\
        10 & 807 & 10 & 2 & 1009 & 93 & 12 & 1x & 93837 & 3.6 & 6.51 & 26 & 314 & 10 & 4 & 389 & 240 & 10 & 3 & 93360 & 8.7 & 3.74 \\
        9 & 947 & 10 & 2 & 1705 & 58 & 7 & 1x & 98890 & 3.6 & 6.38 & 29 & 280 & 10 & 3 & 301 & 310 & 10 & 4 & 93310 & 8.7 & 3.71 \\
        5 & 867 & 10 & 2 & 867 & 58 & 7 & 1x & 50286 & 3.6 & 6.34 & 32 & 260 & 10 & 3 & 333 & 286 & 9 & 4 & 95238 & 8.7 & 3.62 \\
        5 & 847 & 10 & 2 & 847 & 58 & 7 & 1x & 49126 & 3.6 & 6.33 & 31 & 260 & 10 & 3 & 323 & 286 & 9 & 4 & 92378 & 8.7 & 3.62 \\
        12 & 687 & 10 & 2 & 1031 & 93 & 13 & 1 & 95883 & 3.6 & 6.33 & 19 & 214 & 10 & 4 & 226 & 204 & 8 & 3 & 46104 & 8.8 & 3.60 \\
        5 & 747 & 10 & 2 & 747 & 58 & 7 & 1x & 43326 & 3.6 & 6.31 & 27 & 314 & 10 & 4 & 314 & 310 & 10 & 4 & 97340 & 10.0 & 3.45 \\
        5 & 727 & 10 & 2 & 727 & 58 & 7 & 1x & 42166 & 3.6 & 6.31 & 27 & 314 & 10 & 4 & 340 & 286 & 9 & 4 & 97240 & 10.0 & 3.35 \\
        5 & 707 & 10 & 2 & 707 & 58 & 7 & 1x & 41006 & 3.6 & 6.30 & 25 & 347 & 10 & 5 & 347 & 286 & 9 & 4 & 99242 & 14.6 & 3.16 \\
        7 & 507 & 10 & 2 & 507 & 81 & 10 & 1x & 41067 & 3.6 & 6.29 & 35 & 247 & 10 & 5 & 299 & 326 & 9 & 4 & 97474 & 14.8 & 3.13 \\
        5 & 607 & 10 & 2 & 607 & 58 & 7 & 1x & 35206 & 3.6 & 6.27 & 29 & 300 & 10 & 6 & 300 & 326 & 9 & 4 & 97800 & 16.7 & 2.97 \\
        5 & 487 & 10 & 2 & 487 & 58 & 7 & 1x & 28246 & 3.6 & 6.25 & 29 & 294 & 10 & 7 & 294 & 326 & 9 & 4 & 95844 & 18.9 & 2.91 \\
        5 & 467 & 10 & 2 & 467 & 58 & 7 & 1x & 27086 & 3.6 & 6.24 & 33 & 254 & 10 & 7 & 254 & 366 & 9 & 4 & 92964 & 19.1 & 2.90 \\
        6 & 307 & 10 & 2 & 307 & 69 & 8 & 1x & 21183 & 3.6 & 6.15 & 19 & 460 & 10 & 9 & 486 & 204 & 8 & 3 & 99144 & 21.3 & 2.87 \\
        27 & 307 & 10 & 2 & 1382 & 69 & 9 & 1 & 95358 & 3.6 & 6.14 & 25 & 340 & 10 & 9 & 387 & 244 & 8 & 3 & 94428 & 21.8 & 2.84 \\
        6 & 247 & 10 & 2 & 247 & 69 & 8 & 1x & 17043 & 3.6 & 6.14 & 25 & 354 & 10 & 10 & 403 & 244 & 8 & 3 & 98332 & 24.0 & 2.79 \\
        14 & 294 & 10 & 1 & 317 & 152 & 11 & 2 & 48184 & 3.7 & 6.08 & 22 & 387 & 10 & 11 & 387 & 244 & 8 & 3 & 94428 & 26.0 & 2.72 \\
        12 & 194 & 10 & 1 & 233 & 116 & 8 & 2 & 27028 & 3.7 & 6.06 & 17 & 520 & 10 & 12 & 305 & 326 & 9 & 4 & 99430 & 28.8 & 2.71 \\
        125 & 67 & 10 & 2 & 838 & 110 & 9 & 1x & 92180 & 3.8 & 6.00 & 21 & 420 & 10 & 12 & 268 & 366 & 9 & 4 & 98088 & 29.3 & 2.70 \\
        21 & 158 & 8 & 1 & 277 & 112 & 8 & 2 & 31024 & 4.0 & 5.98 & 23 & 380 & 10 & 12 & 237 & 406 & 9 & 4 & 96222 & 29.6 & 2.70 \\
        19 & 66 & 7 & 1 & 66 & 151 & 9 & 2 & 9966 & 4.2 & 5.93 & 16 & 534 & 10 & 13 & 295 & 326 & 9 & 4 & 96170 & 30.9 & 2.67 \\
        21 & 66 & 7 & 1 & 73 & 151 & 9 & 2 & 11023 & 4.2 & 5.93 & 21 & 414 & 10 & 13 & 235 & 406 & 9 & 4 & 95410 & 31.6 & 2.64 \\
        17 & 67 & 8 & 2 & 114 & 88 & 7 & 1x & 10032 & 4.4 & 5.92 & 15 & 587 & 10 & 14 & 304 & 326 & 9 & 4 & 99104 & 32.8 & 2.61 \\
        17 & 67 & 8 & 2 & 127 & 79 & 7 & 1 & 10033 & 4.4 & 5.92 & 19 & 447 & 10 & 14 & 230 & 406 & 9 & 4 & 93380 & 33.7 & 2.60 \\
        26 & 62 & 6 & 1 & 81 & 136 & 8 & 2 & 11016 & 4.5 & 5.90 & 13 & 680 & 10 & 15 & 305 & 326 & 9 & 4 & 99430 & 34.6 & 2.59 \\
        84 & 46 & 6 & 1 & 155 & 168 & 9 & 2 & 26040 & 4.5 & 5.87 & 17 & 520 & 10 & 15 & 268 & 366 & 9 & 4 & 98088 & 35.4 & 2.57 \\
        9 & 880 & 10 & 3 & 1132 & 81 & 10 & 1x & 91692 & 4.8 & 5.77 & 11 & 794 & 10 & 16 & 302 & 326 & 9 & 4 & 98452 & 36.3 & 2.55 \\
        10 & 800 & 10 & 3 & 1143 & 81 & 10 & 1x & 92583 & 4.8 & 5.76 & 16 & 554 & 10 & 16 & 269 & 366 & 9 & 4 & 98454 & 37.5 & 2.54 \\
        11 & 760 & 10 & 3 & 1195 & 81 & 10 & 1x & 96795 & 4.8 & 5.76 & 11 & 787 & 10 & 17 & 299 & 326 & 9 & 4 & 97474 & 38.5 & 2.53 \\
        11 & 740 & 10 & 3 & 1163 & 81 & 10 & 1x & 94203 & 4.8 & 5.76 & 13 & 680 & 10 & 18 & 402 & 244 & 8 & 3 & 98088 & 39.9 & 2.53 \\
        11 & 720 & 10 & 3 & 1132 & 81 & 10 & 1x & 91692 & 4.8 & 5.75 & 9 & 900 & 10 & 18 & 280 & 326 & 9 & 4 & 91280 & 40.2 & 2.51 \\
        13 & 660 & 10 & 3 & 1226 & 81 & 10 & 1x & 99306 & 4.8 & 5.75 & 10 & 820 & 10 & 18 & 283 & 326 & 9 & 4 & 92258 & 40.5 & 2.51 \\
        13 & 640 & 10 & 3 & 1189 & 81 & 10 & 1x & 96309 & 4.8 & 5.75 & 11 & 800 & 10 & 18 & 304 & 326 & 9 & 4 & 99104 & 40.6 & 2.50 \\
        7 & 620 & 10 & 3 & 620 & 81 & 10 & 1x & 50220 & 4.8 & 5.70 & 13 & 680 & 10 & 18 & 268 & 366 & 9 & 4 & 98088 & 41.1 & 2.47 \\
        7 & 580 & 10 & 3 & 580 & 81 & 10 & 1x & 46980 & 4.8 & 5.69 & 13 & 674 & 10 & 19 & 266 & 366 & 9 & 4 & 97356 & 43.4 & 2.47 \\
        23 & 354 & 10 & 1 & 354 & 269 & 13 & 3 & 95226 & 4.9 & 5.63 & 9 & 967 & 10 & 20 & 301 & 326 & 9 & 4 & 98126 & 44.2 & 2.45 \\
        7 & 460 & 10 & 3 & 537 & 69 & 8 & 1x & 37053 & 4.9 & 5.54 & 11 & 747 & 10 & 20 & 249 & 366 & 9 & 4 & 91134 & 45.2 & 2.43 \\
        58 & 74 & 10 & 1 & 159 & 309 & 13 & 3 & 49131 & 4.9 & 5.54 & 14 & 627 & 10 & 17 & 209 & 465 & 9 & 5 & 97185 & 78.5 & 2.41 \\
        39 & 74 & 10 & 1 & 107 & 309 & 13 & 3 & 33063 & 4.9 & 5.51 & 11 & 780 & 10 & 18 & 226 & 425 & 9 & 5 & 96050 & 80.3 & 2.40 \\
        13 & 647 & 10 & 2 & 647 & 152 & 11 & 2 & 98344 & 4.9 & 5.41 & 13 & 660 & 10 & 18 & 205 & 465 & 9 & 5 & 95325 & 80.7 & 2.39 \\
        12 & 707 & 10 & 2 & 707 & 140 & 10 & 2 & 98980 & 4.9 & 5.36 & 13 & 674 & 10 & 19 & 209 & 465 & 9 & 5 & 97185 & 82.9 & 2.38 \\
        12 & 667 & 10 & 2 & 667 & 140 & 10 & 2 & 93380 & 4.9 & 5.35 & 10 & 867 & 10 & 20 & 229 & 425 & 9 & 5 & 97325 & 84.4 & 2.35 \\
        13 & 627 & 10 & 2 & 680 & 140 & 10 & 2 & 95200 & 4.9 & 5.35 & 11 & 800 & 10 & 18 & 205 & 484 & 9 & 6 & 99220 & 89.7 & 2.35 \\
        13 & 607 & 10 & 2 & 658 & 140 & 10 & 2 & 92120 & 4.9 & 5.35 & 10 & 854 & 10 & 19 & 199 & 484 & 9 & 6 & 96316 & 91.6 & 2.34 \\
        15 & 527 & 10 & 2 & 659 & 140 & 10 & 2 & 92260 & 4.9 & 5.33 & 9 & 967 & 10 & 20 & 203 & 484 & 9 & 6 & 98252 & 93.2 & 2.30 \\
        15 & 527 & 10 & 2 & 719 & 128 & 9 & 2 & 92032 & 4.9 & 5.27 & 10 & 874 & 10 & 19 & 117 & 840 & 10 & 9 & 98280 & 117.2 & 2.29 \\
        17 & 467 & 10 & 2 & 722 & 128 & 9 & 2 & 92416 & 4.9 & 5.27 & 10 & 880 & 10 & 18 & 109 & 910 & 10 & 10 & 99190 & 124.0 & 2.29 \\
        19 & 427 & 10 & 2 & 738 & 128 & 9 & 2 & 94464 & 4.9 & 5.26 & 9 & 927 & 10 & 20 & 103 & 910 & 10 & 10 & 93730 & 128.0 & 2.27 \\
        \hline
\end{tabular}

\vspace{0.1cm}
There are 3,985,168 data points with 162 of them on the Pareto frontier.
\label{tab:100k-rate085}
\end{table*}

\begin{table*}
\setlength{\tabcolsep}{2pt}
\centering
\caption{Pareto-Efficient Code Parameters With Latency $\leq 10,000$ Bits and Rate $\approx 0.937$}
\begin{tabular}{|c|c|c|c|c|c|c|c|c|c|c||c|c|c|c|c|c|c|c|c|c|c|c|c|}
    \hline
        $M$ & $N$ & $B$ & $T$ & $m$ & $n$ & $b$ & $t$ & Latency & Complexity & Gap (dB) & $M$ & $N$ & $B$ & $T$ & $m$ & $n$ & $b$ & $t$ & Latency & Complexity & Gap (dB)\\\hline
        --- & --- & --- & --- & 1 & 223 & 14 & 1 & 223 & 1.1 & 7.65 & 2 & 375 & 9 & 5 & 21 & 336 & 11 & 1x & 7056 & 11.7 & 4.61 \\
        --- & --- & --- & --- & 1 & 207 & 13 & 1 & 207 & 1.1 & 7.63 & 1 & 851 & 10 & 6 & 39 & 232 & 11 & 1x & 9048 & 12.2 & 4.53 \\
        --- & --- & --- & --- & 1 & 191 & 12 & 1 & 191 & 1.1 & 7.61 & 1 & 671 & 10 & 6 & 24 & 293 & 12 & 1x & 7032 & 12.3 & 4.46 \\
        --- & --- & --- & --- & 1 & 175 & 11 & 1 & 175 & 1.1 & 7.59 & 1 & 671 & 10 & 6 & 27 & 261 & 10 & 1x & 7047 & 12.3 & 4.41 \\
        --- & --- & --- & --- & 1 & 159 & 10 & 1 & 159 & 1.1 & 7.57 & 2 & 431 & 10 & 6 & 29 & 312 & 11 & 1x & 9048 & 12.6 & 4.34 \\
        --- & --- & --- & --- & 1 & 143 & 9 & 1 & 143 & 1.1 & 7.54 & 2 & 391 & 10 & 6 & 17 & 472 & 11 & 1x & 8024 & 12.6 & 4.23 \\
        --- & --- & --- & --- & 1 & 127 & 8 & 1 & 127 & 1.1 & 7.51 & 1 & 863 & 10 & 7 & 29 & 312 & 11 & 1x & 9048 & 14.1 & 4.19 \\
        --- & --- & --- & --- & 1 & 112 & 7 & 1 & 112 & 1.1 & 7.47 & 1 & 223 & 10 & 7 & --- & --- & --- & --- & 2230 & 14.4 & 4.16 \\
        --- & --- & --- & --- & 1 & 445 & 14 & 2 & 445 & 2.2 & 6.37 & 2 & 483 & 10 & 7 & 21 & 472 & 11 & 1x & 9912 & 14.5 & 4.01 \\
        --- & --- & --- & --- & 1 & 413 & 13 & 2 & 413 & 2.2 & 6.35 & 1 & 774 & 10 & 8 & 21 & 384 & 13 & 1x & 8064 & 16.1 & 3.92 \\
        --- & --- & --- & --- & 1 & 381 & 12 & 2 & 381 & 2.2 & 6.33 & 1 & 674 & 10 & 8 & 17 & 413 & 12 & 1x & 7021 & 16.2 & 3.88 \\
        --- & --- & --- & --- & 1 & 350 & 11 & 2 & 350 & 2.2 & 6.29 & 2 & 394 & 10 & 8 & 9 & 893 & 12 & 1x & 8037 & 16.8 & 3.86 \\
        --- & --- & --- & --- & 1 & 318 & 10 & 2 & 318 & 2.2 & 6.26 & 2 & 483 & 10 & 7 & 7 & 1424 & 11 & 4 & 9968 & 17.7 & 3.66 \\
        --- & --- & --- & --- & 1 & 286 & 9 & 2 & 286 & 2.2 & 6.23 & 1 & 586 & 10 & 9 & 12 & 504 & 13 & 1x & 6048 & 18.4 & 3.57 \\
        --- & --- & --- & --- & 1 & 254 & 8 & 2 & 254 & 2.2 & 6.19 & 1 & 870 & 10 & 11 & 23 & 392 & 11 & 1x & 9016 & 22.0 & 3.48 \\
        --- & --- & --- & --- & 1 & 667 & 14 & 3 & 667 & 3.2 & 5.56 & 1 & 961 & 10 & 12 & 31 & 320 & 9 & 1x & 9920 & 23.8 & 3.33 \\
        --- & --- & --- & --- & 1 & 620 & 13 & 3 & 620 & 3.2 & 5.53 & 1 & 881 & 10 & 12 & 21 & 432 & 11 & 1x & 9072 & 23.9 & 3.33 \\
        --- & --- & --- & --- & 1 & 572 & 12 & 3 & 572 & 3.2 & 5.50 & 1 & 873 & 10 & 13 & 23 & 392 & 11 & 1x & 9016 & 26.0 & 3.32 \\
        --- & --- & --- & --- & 1 & 524 & 11 & 3 & 524 & 3.3 & 5.46 & 1 & 873 & 10 & 13 & 20 & 453 & 12 & 1x & 9060 & 26.0 & 3.29 \\
        --- & --- & --- & --- & 1 & 477 & 10 & 3 & 477 & 3.3 & 5.43 & 1 & 685 & 10 & 14 & 10 & 704 & 13 & 1x & 7040 & 28.6 & 3.28 \\
        --- & --- & --- & --- & 1 & 429 & 9 & 3 & 429 & 3.3 & 5.39 & 1 & 445 & 10 & 14 & --- & --- & --- & --- & 4450 & 28.8 & 3.20 \\
        --- & --- & --- & --- & 1 & 889 & 14 & 4 & 889 & 4.4 & 4.99 & 1 & 877 & 10 & 15 & 18 & 501 & 10 & 1x & 9018 & 30.1 & 3.16 \\
        --- & --- & --- & --- & 1 & 826 & 13 & 4 & 826 & 4.4 & 4.96 & 1 & 877 & 10 & 15 & 17 & 533 & 12 & 1x & 9061 & 30.1 & 3.14 \\
        --- & --- & --- & --- & 1 & 762 & 12 & 4 & 762 & 4.4 & 4.93 & 1 & 477 & 10 & 15 & --- & --- & --- & --- & 4770 & 30.9 & 3.12 \\
        --- & --- & --- & --- & 1 & 699 & 11 & 4 & 699 & 4.4 & 4.89 & 1 & 597 & 10 & 15 & 2 & 3013 & 13 & 1 & 6026 & 31.1 & 3.11 \\
        --- & --- & --- & --- & 1 & 635 & 10 & 4 & 635 & 4.4 & 4.84 & 1 & 968 & 10 & 16 & 22 & 453 & 12 & 1x & 9966 & 32.0 & 3.09 \\
        3 & 325 & 9 & 4 & 25 & 362 & 10 & 1x & 9050 & 6.6 & 4.78 & 1 & 788 & 10 & 16 & 13 & 621 & 10 & 1x & 8073 & 32.5 & 2.96 \\
        1 & 159 & 10 & 5 & --- & --- & --- & --- & 1590 & 10.3 & 4.73 & 1 & 972 & 10 & 18 & 18 & 552 & 11 & 1x & 9936 & 36.1 & 2.96 \\
        3 & 259 & 10 & 5 & 14 & 573 & 12 & 1x & 8022 & 10.9 & 4.70 & 1 & 572 & 10 & 18 & --- & --- & --- & --- & 5720 & 37.0 & 2.93 \\
        1 & 159 & 9 & 5 & --- & --- & --- & --- & 1431 & 11.5 & 4.68 & 1 & 692 & 10 & 18 & 1 & 6933 & 13 & 1 & 6933 & 37.2 & 2.82 \\
        2 & 483 & 9 & 5 & 27 & 336 & 11 & 1x & 9072 & 11.5 & 4.64 & 1 & 635 & 10 & 20 & --- & --- & --- & --- & 6350 & 41.1 & 2.82 \\
        2 & 375 & 9 & 5 & 15 & 463 & 12 & 1x & 6945 & 11.7 & 4.62 & --- & --- & --- & --- & 1 & 2858 & 12 & 15 & 2858 & 116.2 & 2.78 \\
        \hline
\end{tabular}

\vspace{0.1cm}
There are 8,096 data points with 64 of them on the Pareto frontier.
\label{tab:10k-rate0937}
\end{table*}

\begin{table*}
\setlength{\tabcolsep}{2pt}
\centering
\caption{Pareto-Efficient Code Parameters With Latency $\leq 30,000$ Bits and Rate $\approx 0.937$}
\begin{tabular}{|c|c|c|c|c|c|c|c|c|c|c||c|c|c|c|c|c|c|c|c|c|c|c|c|}
    \hline
        $M$ & $N$ & $B$ & $T$ & $m$ & $n$ & $b$ & $t$ & Latency & Complexity & Gap (dB) & $M$ & $N$ & $B$ & $T$ & $m$ & $n$ & $b$ & $t$ & Latency & Complexity & Gap (dB)\\\hline
        --- & --- & --- & --- & 1 & 223 & 14 & 1 & 223 & 1.1 & 7.65 & 9 & 279 & 10 & 5 & 74 & 352 & 11 & 1x & 26048 & 10.8 & 4.23 \\
        --- & --- & --- & --- & 1 & 207 & 13 & 1 & 207 & 1.1 & 7.63 & 9 & 357 & 9 & 5 & 119 & 252 & 8 & 1x & 29988 & 11.7 & 4.22 \\
        --- & --- & --- & --- & 1 & 191 & 12 & 1 & 191 & 1.1 & 7.61 & 3 & 931 & 10 & 6 & 112 & 261 & 10 & 1x & 29232 & 12.2 & 4.10 \\
        --- & --- & --- & --- & 1 & 175 & 11 & 1 & 175 & 1.1 & 7.59 & 3 & 931 & 10 & 6 & 122 & 240 & 9 & 1x & 29280 & 12.2 & 4.08 \\
        --- & --- & --- & --- & 1 & 159 & 10 & 1 & 159 & 1.1 & 7.57 & 4 & 691 & 10 & 6 & 93 & 312 & 11 & 1x & 29016 & 12.3 & 4.03 \\
        --- & --- & --- & --- & 1 & 143 & 9 & 1 & 143 & 1.1 & 7.54 & 4 & 651 & 10 & 6 & 119 & 229 & 8 & 1x & 27251 & 12.3 & 4.02 \\
        --- & --- & --- & --- & 1 & 127 & 8 & 1 & 127 & 1.1 & 7.51 & 4 & 531 & 10 & 6 & 65 & 341 & 10 & 1x & 22165 & 12.4 & 3.99 \\
        --- & --- & --- & --- & 1 & 112 & 7 & 1 & 112 & 1.1 & 7.47 & 6 & 451 & 10 & 6 & 65 & 432 & 11 & 1x & 28080 & 12.5 & 3.91 \\
        --- & --- & --- & --- & 1 & 445 & 14 & 2 & 445 & 2.2 & 6.37 & 7 & 391 & 10 & 6 & 61 & 461 & 10 & 1x & 28121 & 12.6 & 3.89 \\
        --- & --- & --- & --- & 1 & 413 & 13 & 2 & 413 & 2.2 & 6.35 & 10 & 271 & 10 & 6 & 57 & 493 & 12 & 1x & 28101 & 13.0 & 3.86 \\
        --- & --- & --- & --- & 1 & 381 & 12 & 2 & 381 & 2.2 & 6.33 & 8 & 391 & 9 & 6 & 67 & 434 & 10 & 1x & 29078 & 13.9 & 3.85 \\
        --- & --- & --- & --- & 1 & 350 & 11 & 2 & 350 & 2.2 & 6.29 & 8 & 391 & 9 & 6 & 77 & 379 & 9 & 1x & 29183 & 13.9 & 3.85 \\
        --- & --- & --- & --- & 1 & 318 & 10 & 2 & 318 & 2.2 & 6.26 & 4 & 703 & 10 & 7 & 97 & 301 & 10 & 1x & 29197 & 14.2 & 3.79 \\
        --- & --- & --- & --- & 1 & 286 & 9 & 2 & 286 & 2.2 & 6.23 & 3 & 703 & 10 & 7 & 59 & 373 & 12 & 1x & 22007 & 14.2 & 3.71 \\
        --- & --- & --- & --- & 1 & 254 & 8 & 2 & 254 & 2.2 & 6.19 & 5 & 543 & 10 & 7 & 74 & 381 & 10 & 1x & 28194 & 14.4 & 3.70 \\
        17 & 152 & 10 & 1 & 162 & 168 & 8 & 1 & 27216 & 2.3 & 5.93 & 7 & 403 & 10 & 7 & 66 & 440 & 9 & 1x & 29040 & 14.7 & 3.69 \\
        23 & 112 & 10 & 1 & 112 & 240 & 9 & 1x & 26880 & 2.3 & 5.88 & 4 & 671 & 10 & 6 & 21 & 1336 & 14 & 4 & 28056 & 15.5 & 3.68 \\
        27 & 92 & 10 & 1 & 92 & 280 & 9 & 1x & 25760 & 2.3 & 5.83 & 4 & 691 & 10 & 6 & 33 & 880 & 10 & 4 & 29040 & 15.6 & 3.53 \\
        23 & 112 & 10 & 1 & 118 & 229 & 8 & 1x & 27022 & 2.3 & 5.78 & 4 & 654 & 10 & 8 & 85 & 320 & 9 & 1x & 27200 & 16.2 & 3.52 \\
        23 & 152 & 8 & 1 & 152 & 192 & 8 & 1 & 29184 & 2.6 & 5.76 & 6 & 474 & 10 & 8 & 73 & 400 & 9 & 1x & 29200 & 16.5 & 3.49 \\
        29 & 112 & 8 & 1 & 112 & 241 & 8 & 1x & 26992 & 2.6 & 5.75 & 6 & 474 & 10 & 8 & 53 & 552 & 11 & 1x & 29256 & 16.5 & 3.49 \\
        33 & 108 & 7 & 1 & 112 & 233 & 8 & 1x & 26096 & 2.8 & 5.75 & 6 & 474 & 10 & 8 & 61 & 480 & 9 & 1x & 29280 & 16.6 & 3.47 \\
        79 & 58 & 6 & 1 & 65 & 435 & 9 & 1 & 28275 & 3.2 & 5.69 & 8 & 391 & 9 & 6 & 19 & 1529 & 11 & 4 & 29051 & 17.1 & 3.45 \\
        --- & --- & --- & --- & 1 & 667 & 14 & 3 & 667 & 3.2 & 5.56 & 3 & 703 & 10 & 7 & 23 & 960 & 10 & 4 & 22080 & 17.5 & 3.44 \\
        --- & --- & --- & --- & 1 & 620 & 13 & 3 & 620 & 3.2 & 5.53 & 5 & 503 & 10 & 7 & 18 & 1448 & 12 & 4 & 26064 & 17.7 & 3.43 \\
        --- & --- & --- & --- & 1 & 572 & 12 & 3 & 572 & 3.2 & 5.50 & 6 & 483 & 10 & 7 & 23 & 1304 & 11 & 4 & 29992 & 17.7 & 3.36 \\
        --- & --- & --- & --- & 1 & 524 & 11 & 3 & 524 & 3.3 & 5.46 & 4 & 606 & 10 & 9 & 61 & 413 & 12 & 1x & 25193 & 18.3 & 3.21 \\
        --- & --- & --- & --- & 1 & 477 & 10 & 3 & 477 & 3.3 & 5.43 & 7 & 394 & 9 & 9 & 30 & 840 & 11 & 1x & 25200 & 20.9 & 3.19 \\
        --- & --- & --- & --- & 1 & 429 & 9 & 3 & 429 & 3.3 & 5.39 & 3 & 930 & 10 & 11 & 104 & 280 & 9 & 1x & 29120 & 21.9 & 3.18 \\
        21 & 124 & 10 & 2 & 131 & 208 & 8 & 1 & 27248 & 3.6 & 5.04 & 3 & 870 & 10 & 11 & 85 & 320 & 9 & 1x & 27200 & 22.0 & 3.14 \\
        25 & 124 & 9 & 2 & 141 & 206 & 8 & 1 & 29046 & 3.9 & 5.04 & 3 & 810 & 10 & 11 & 70 & 360 & 9 & 1x & 25200 & 22.1 & 3.14 \\
        19 & 176 & 8 & 2 & 176 & 160 & 8 & 1 & 28160 & 4.2 & 5.04 & 4 & 710 & 10 & 11 & 57 & 512 & 11 & 1x & 29184 & 22.2 & 3.10 \\
        23 & 152 & 8 & 2 & 152 & 192 & 8 & 1 & 29184 & 4.2 & 4.96 & 4 & 710 & 10 & 11 & 73 & 400 & 9 & 1x & 29200 & 22.2 & 3.09 \\
        --- & --- & --- & --- & 1 & 762 & 12 & 4 & 762 & 4.4 & 4.93 & 5 & 550 & 10 & 11 & 43 & 653 & 12 & 1x & 28079 & 22.6 & 3.00 \\
        --- & --- & --- & --- & 1 & 699 & 11 & 4 & 699 & 4.4 & 4.89 & 3 & 770 & 10 & 11 & 19 & 1264 & 11 & 4 & 24016 & 25.4 & 2.94 \\
        --- & --- & --- & --- & 1 & 635 & 10 & 4 & 635 & 4.4 & 4.84 & 3 & 913 & 10 & 13 & 67 & 421 & 10 & 1x & 28207 & 25.9 & 2.93 \\
        3 & 767 & 10 & 4 & 89 & 272 & 11 & 1x & 24208 & 6.0 & 4.83 & 3 & 841 & 10 & 12 & 23 & 1136 & 12 & 3 & 26128 & 26.2 & 2.92 \\
        4 & 667 & 10 & 4 & 103 & 272 & 11 & 1x & 28016 & 6.0 & 4.78 & 4 & 681 & 10 & 12 & 21 & 1336 & 12 & 3 & 28056 & 26.5 & 2.92 \\
        4 & 527 & 10 & 4 & 85 & 261 & 10 & 1x & 22185 & 6.0 & 4.77 & 3 & 841 & 10 & 12 & 16 & 1632 & 13 & 4 & 26112 & 27.3 & 2.88 \\
        6 & 467 & 10 & 4 & 85 & 344 & 13 & 1x & 29240 & 6.0 & 4.74 & 4 & 701 & 10 & 12 & 19 & 1528 & 12 & 4 & 29032 & 27.6 & 2.86 \\
        6 & 447 & 10 & 4 & 73 & 384 & 13 & 1x & 28032 & 6.0 & 4.67 & 4 & 701 & 10 & 12 & 21 & 1384 & 11 & 4 & 29064 & 27.6 & 2.81 \\
        7 & 387 & 10 & 4 & 80 & 352 & 11 & 1x & 28160 & 6.0 & 4.61 & 3 & 933 & 10 & 13 & 19 & 1528 & 12 & 4 & 29032 & 29.1 & 2.75 \\
        7 & 387 & 10 & 4 & 101 & 280 & 9 & 1x & 28280 & 6.1 & 4.61 & 3 & 933 & 10 & 13 & 23 & 1264 & 11 & 4 & 29072 & 29.2 & 2.73 \\
        8 & 327 & 10 & 4 & 69 & 392 & 11 & 1x & 27048 & 6.1 & 4.60 & 3 & 905 & 10 & 14 & 17 & 1648 & 12 & 4 & 28016 & 31.2 & 2.71 \\
        8 & 327 & 10 & 4 & 85 & 320 & 9 & 1x & 27200 & 6.1 & 4.58 & 3 & 937 & 10 & 15 & 19 & 1528 & 12 & 4 & 29032 & 33.2 & 2.67 \\
        11 & 247 & 10 & 4 & 56 & 501 & 10 & 1x & 28056 & 6.2 & 4.57 & 4 & 688 & 10 & 16 & 17 & 1653 & 11 & 3 & 28101 & 35.0 & 2.67 \\
        8 & 387 & 9 & 4 & 89 & 326 & 10 & 1x & 29014 & 6.6 & 4.55 & 3 & 848 & 10 & 16 & 14 & 1864 & 11 & 4 & 26096 & 35.5 & 2.64 \\
        9 & 347 & 9 & 4 & 95 & 307 & 9 & 1x & 29165 & 6.6 & 4.53 & 3 & 912 & 10 & 18 & 14 & 2008 & 12 & 4 & 28112 & 39.5 & 2.61 \\
        15 & 207 & 9 & 4 & 56 & 518 & 13 & 1x & 29008 & 6.8 & 4.49 & 3 & 944 & 10 & 19 & 20 & 1456 & 12 & 3 & 29120 & 40.5 & 2.61 \\
        17 & 187 & 9 & 4 & 53 & 552 & 11 & 1x & 29256 & 6.8 & 4.47 & 3 & 952 & 10 & 16 & 82 & 360 & 10 & 1x & 29520 & 53.8 & 2.54 \\
        16 & 215 & 8 & 4 & 35 & 806 & 13 & 1x & 28210 & 7.5 & 4.42 & 3 & 954 & 10 & 17 & 64 & 461 & 11 & 1x & 29504 & 58.0 & 2.50 \\
        8 & 387 & 9 & 4 & 25 & 1164 & 12 & 4 & 29100 & 9.9 & 4.37 & 3 & 902 & 10 & 19 & 70 & 400 & 10 & 1x & 28000 & 60.5 & 2.37 \\
        5 & 559 & 10 & 5 & 97 & 301 & 10 & 1x & 29197 & 10.4 & 4.37 & 3 & 960 & 10 & 20 & 67 & 440 & 10 & 1x & 29480 & 62.2 & 2.31 \\
        7 & 399 & 10 & 5 & 74 & 392 & 11 & 1x & 29008 & 10.6 & 4.30 & 3 & 928 & 10 & 16 & 19 & 1544 & 11 & 4 & 29336 & 124.3 & 2.27 \\
        8 & 339 & 10 & 5 & 57 & 493 & 12 & 1x & 28101 & 10.7 & 4.30 & & & & & & & & & & & \\
        \hline
\end{tabular}

\vspace{0.1cm}
There are 366,552 data points with 109 of them on the Pareto frontier.
\label{tab:30k-rate0937}
\end{table*}

\begin{table*}
\setlength{\tabcolsep}{2pt}
\centering
\caption{Pareto-Efficient Code Parameters With Latency $\leq 100,000$ Bits and Rate $\approx 0.937$}
\begin{tabular}{|c|c|c|c|c|c|c|c|c|c|c||c|c|c|c|c|c|c|c|c|c|c|c|c|}
    \hline
        $M$ & $N$ & $B$ & $T$ & $m$ & $n$ & $b$ & $t$ & Latency & Complexity & Gap (dB) & $M$ & $N$ & $B$ & $T$ & $m$ & $n$ & $b$ & $t$ & Latency & Complexity & Gap (dB)\\\hline
        --- & --- & --- & --- & 1 & 223 & 14 & 1 & 223 & 1.1 & 7.65 & 43 & 216 & 10 & 3 & 216 & 448 & 9 & 2 & 96768 & 5.9 & 3.63 \\
        --- & --- & --- & --- & 1 & 207 & 13 & 1 & 207 & 1.1 & 7.63 & 47 & 222 & 9 & 3 & 227 & 432 & 9 & 2 & 98064 & 6.4 & 3.62 \\
        --- & --- & --- & --- & 1 & 191 & 12 & 1 & 191 & 1.1 & 7.61 & 29 & 361 & 9 & 4 & 361 & 271 & 9 & 1x & 97831 & 6.6 & 3.59 \\
        --- & --- & --- & --- & 1 & 175 & 11 & 1 & 175 & 1.1 & 7.59 & 30 & 316 & 10 & 3 & 113 & 880 & 10 & 4 & 99440 & 8.1 & 3.58 \\
        --- & --- & --- & --- & 1 & 159 & 10 & 1 & 159 & 1.1 & 7.57 & 27 & 347 & 10 & 4 & 127 & 773 & 11 & 3 & 98171 & 8.3 & 3.58 \\
        --- & --- & --- & --- & 1 & 143 & 9 & 1 & 143 & 1.1 & 7.54 & 33 & 312 & 9 & 3 & 104 & 931 & 10 & 4 & 96824 & 8.5 & 3.54 \\
        --- & --- & --- & --- & 1 & 127 & 8 & 1 & 127 & 1.1 & 7.51 & 29 & 325 & 9 & 4 & 115 & 768 & 10 & 3 & 88320 & 8.8 & 3.53 \\
        --- & --- & --- & --- & 1 & 112 & 7 & 1 & 112 & 1.1 & 7.47 & 47 & 240 & 8 & 3 & 95 & 992 & 10 & 4 & 94240 & 9.0 & 3.51 \\
        --- & --- & --- & --- & 1 & 445 & 14 & 2 & 445 & 2.2 & 6.37 & 22 & 427 & 10 & 4 & 107 & 920 & 10 & 4 & 98440 & 9.3 & 3.40 \\
        --- & --- & --- & --- & 1 & 413 & 13 & 2 & 413 & 2.2 & 6.35 & 20 & 467 & 10 & 4 & 117 & 840 & 10 & 4 & 98280 & 9.3 & 3.26 \\
        --- & --- & --- & --- & 1 & 381 & 12 & 2 & 381 & 2.2 & 6.33 & 17 & 559 & 10 & 5 & 108 & 920 & 10 & 4 & 99360 & 13.7 & 3.26 \\
        --- & --- & --- & --- & 1 & 350 & 11 & 2 & 350 & 2.2 & 6.29 & 19 & 499 & 10 & 5 & 108 & 920 & 10 & 4 & 99360 & 13.8 & 3.25 \\
        --- & --- & --- & --- & 1 & 318 & 10 & 2 & 318 & 2.2 & 6.26 & 19 & 479 & 10 & 5 & 99 & 960 & 10 & 4 & 95040 & 13.8 & 3.25 \\
        --- & --- & --- & --- & 1 & 286 & 9 & 2 & 286 & 2.2 & 6.23 & 23 & 461 & 9 & 6 & 461 & 216 & 8 & 1x & 99576 & 13.8 & 3.18 \\
        --- & --- & --- & --- & 1 & 254 & 8 & 2 & 254 & 2.2 & 6.19 & 17 & 563 & 10 & 7 & 157 & 632 & 11 & 2 & 99224 & 15.4 & 3.16 \\
        19 & 492 & 10 & 1 & 492 & 200 & 9 & 1x & 98400 & 2.3 & 5.96 & 12 & 791 & 10 & 6 & 113 & 880 & 10 & 4 & 99440 & 15.5 & 3.12 \\
        19 & 472 & 10 & 1 & 472 & 200 & 9 & 1x & 94400 & 2.3 & 5.96 & 13 & 731 & 10 & 6 & 108 & 920 & 10 & 4 & 99360 & 15.5 & 3.09 \\
        19 & 452 & 10 & 1 & 452 & 200 & 9 & 1x & 90400 & 2.3 & 5.95 & 14 & 671 & 10 & 6 & 107 & 920 & 10 & 4 & 98440 & 15.6 & 3.09 \\
        19 & 432 & 10 & 1 & 432 & 200 & 9 & 1x & 86400 & 2.3 & 5.95 & 16 & 591 & 10 & 6 & 113 & 880 & 10 & 4 & 99440 & 15.7 & 3.08 \\
        19 & 412 & 10 & 1 & 412 & 200 & 9 & 1x & 82400 & 2.3 & 5.95 & 21 & 451 & 10 & 6 & 73 & 1344 & 11 & 4 & 98112 & 15.8 & 3.07 \\
        19 & 392 & 10 & 1 & 392 & 200 & 9 & 1x & 78400 & 2.3 & 5.85 & 22 & 431 & 10 & 6 & 71 & 1384 & 11 & 4 & 98264 & 15.8 & 3.07 \\
        29 & 292 & 10 & 1 & 530 & 168 & 8 & 1 & 89040 & 2.3 & 5.83 & 27 & 351 & 10 & 6 & 68 & 1448 & 12 & 4 & 98464 & 16.0 & 3.05 \\
        29 & 112 & 10 & 1 & 112 & 301 & 10 & 1x & 33712 & 2.3 & 5.81 & 26 & 385 & 9 & 7 & 371 & 252 & 8 & 1x & 93492 & 16.2 & 3.00 \\
        37 & 132 & 10 & 1 & 213 & 240 & 9 & 1x & 51120 & 2.3 & 5.74 & 19 & 503 & 10 & 7 & 98 & 1013 & 11 & 3 & 99274 & 16.6 & 2.99 \\
        44 & 86 & 9 & 1 & 95 & 372 & 11 & 1x & 35340 & 2.4 & 5.70 & 19 & 503 & 10 & 7 & 102 & 973 & 11 & 3 & 99246 & 16.6 & 2.99 \\
        37 & 152 & 8 & 1 & 245 & 192 & 8 & 1 & 47040 & 2.6 & 5.69 & 19 & 483 & 10 & 7 & 86 & 1109 & 13 & 3 & 95374 & 16.7 & 2.82 \\
        43 & 112 & 8 & 1 & 124 & 324 & 11 & 1x & 40176 & 2.6 & 5.69 & 25 & 374 & 10 & 8 & 48 & 1989 & 13 & 3 & 95472 & 19.0 & 2.78 \\
        60 & 124 & 7 & 1 & 233 & 233 & 9 & 1 & 54289 & 2.8 & 5.69 & 9 & 954 & 10 & 8 & 98 & 920 & 10 & 4 & 90160 & 19.3 & 2.78 \\
        79 & 88 & 7 & 1 & 145 & 347 & 10 & 1x & 50315 & 2.8 & 5.67 & 10 & 934 & 10 & 8 & 117 & 840 & 10 & 4 & 98280 & 19.3 & 2.77 \\
        --- & --- & --- & --- & 1 & 667 & 14 & 3 & 667 & 3.2 & 5.56 & 10 & 854 & 10 & 8 & 93 & 960 & 10 & 4 & 89280 & 19.3 & 2.74 \\
        --- & --- & --- & --- & 1 & 620 & 13 & 3 & 620 & 3.2 & 5.53 & 11 & 854 & 10 & 8 & 107 & 920 & 10 & 4 & 98440 & 19.3 & 2.72 \\
        --- & --- & --- & --- & 1 & 572 & 12 & 3 & 572 & 3.2 & 5.50 & 16 & 594 & 10 & 8 & 99 & 1000 & 10 & 4 & 99000 & 19.6 & 2.68 \\
        --- & --- & --- & --- & 1 & 524 & 11 & 3 & 524 & 3.3 & 5.46 & 15 & 594 & 10 & 8 & 97 & 960 & 10 & 4 & 93120 & 19.6 & 2.64 \\
        --- & --- & --- & --- & 1 & 477 & 10 & 3 & 477 & 3.3 & 5.43 & 13 & 646 & 10 & 9 & 86 & 1016 & 12 & 3 & 87376 & 20.4 & 2.59 \\
        --- & --- & --- & --- & 1 & 429 & 9 & 3 & 429 & 3.3 & 5.39 & 15 & 626 & 10 & 9 & 89 & 1104 & 11 & 4 & 98256 & 21.6 & 2.58 \\
        51 & 172 & 10 & 1 & 172 & 536 & 13 & 2 & 92192 & 3.4 & 4.97 & 16 & 578 & 10 & 10 & 63 & 1509 & 13 & 3 & 95067 & 22.6 & 2.52 \\
        41 & 212 & 10 & 1 & 212 & 432 & 11 & 2 & 91584 & 3.4 & 4.93 & 9 & 998 & 10 & 10 & 98 & 960 & 10 & 4 & 94080 & 23.1 & 2.49 \\
        63 & 132 & 10 & 1 & 160 & 544 & 12 & 2 & 87040 & 3.4 & 4.93 & 10 & 938 & 10 & 10 & 107 & 920 & 10 & 4 & 98440 & 23.2 & 2.49 \\
        45 & 152 & 10 & 1 & 152 & 472 & 11 & 2 & 71744 & 3.4 & 4.90 & 9 & 990 & 10 & 11 & 91 & 1024 & 11 & 4 & 93184 & 25.1 & 2.49 \\
        39 & 212 & 10 & 1 & 218 & 400 & 10 & 2 & 87200 & 3.4 & 4.89 & 9 & 990 & 10 & 11 & 97 & 960 & 10 & 4 & 93120 & 25.1 & 2.41 \\
        35 & 192 & 10 & 1 & 192 & 368 & 9 & 2 & 70656 & 3.4 & 4.85 & 10 & 910 & 10 & 11 & 99 & 960 & 10 & 4 & 95040 & 25.2 & 2.39 \\
        23 & 304 & 10 & 2 & 304 & 240 & 9 & 1x & 72960 & 3.5 & 4.83 & 9 & 1001 & 10 & 12 & 98 & 960 & 10 & 4 & 94080 & 27.1 & 2.34 \\
        30 & 304 & 10 & 2 & 397 & 240 & 10 & 1 & 95280 & 3.5 & 4.75 & 9 & 993 & 10 & 13 & 76 & 1224 & 11 & 4 & 93024 & 29.1 & 2.32 \\
        42 & 224 & 10 & 2 & 325 & 301 & 11 & 1 & 97825 & 3.5 & 4.73 & 10 & 953 & 10 & 13 & 81 & 1224 & 11 & 4 & 99144 & 29.1 & 2.31 \\
        73 & 152 & 8 & 1 & 222 & 420 & 10 & 2 & 93240 & 3.7 & 4.72 & 9 & 1005 & 10 & 14 & 77 & 1224 & 11 & 4 & 94248 & 31.1 & 2.28 \\
        27 & 208 & 9 & 2 & 208 & 252 & 8 & 1x & 52416 & 3.8 & 4.56 & 8 & 945 & 10 & 14 & 55 & 1424 & 11 & 4 & 78320 & 31.2 & 2.27 \\
        99 & 92 & 10 & 1 & 92 & 1029 & 13 & 3 & 94668 & 4.4 & 4.36 & 8 & 1017 & 10 & 15 & 59 & 1424 & 11 & 4 & 84016 & 33.1 & 2.26 \\
        65 & 132 & 10 & 1 & 132 & 680 & 10 & 3 & 89760 & 4.5 & 4.22 & 9 & 1017 & 10 & 15 & 73 & 1304 & 11 & 4 & 95192 & 33.1 & 2.24 \\
        75 & 122 & 9 & 1 & 124 & 696 & 10 & 3 & 86304 & 4.6 & 4.22 & 9 & 1008 & 10 & 16 & 59 & 1584 & 11 & 4 & 93456 & 35.1 & 2.21 \\
        43 & 204 & 10 & 2 & 204 & 448 & 9 & 2 & 91392 & 4.6 & 4.00 & 9 & 860 & 10 & 17 & 61 & 1296 & 13 & 2 & 79056 & 35.4 & 2.19 \\
        52 & 224 & 8 & 2 & 233 & 418 & 9 & 2 & 97394 & 5.2 & 3.99 & 9 & 952 & 10 & 18 & 68 & 1296 & 12 & 3 & 88128 & 38.3 & 2.10 \\
        55 & 208 & 8 & 2 & 212 & 450 & 9 & 2 & 95400 & 5.2 & 3.97 & 7 & 1020 & 10 & 18 & 74 & 1000 & 10 & 3 & 74000 & 97.5 & 2.09 \\
        67 & 176 & 8 & 2 & 204 & 482 & 9 & 2 & 98328 & 5.3 & 3.84 & 9 & 994 & 10 & 17 & 82 & 1133 & 11 & 3 & 92906 & 101.8 & 2.05 \\
        99 & 104 & 9 & 1 & 104 & 931 & 10 & 4 & 96824 & 5.7 & 3.84 & 9 & 1022 & 10 & 19 & 71 & 1333 & 11 & 3 & 94643 & 105.8 & 2.01 \\
        113 & 104 & 8 & 1 & 106 & 928 & 10 & 4 & 98368 & 5.9 & 3.83 & 9 & 1006 & 10 & 15 & 68 & 1384 & 11 & 4 & 94112 & 120.1 & 2.00 \\
        114 & 104 & 8 & 1 & 107 & 928 & 10 & 4 & 99296 & 5.9 & 3.83 & 10 & 952 & 10 & 16 & 61 & 1624 & 11 & 4 & 99064 & 123.2 & 1.98 \\
        121 & 96 & 8 & 1 & 98 & 992 & 10 & 4 & 97216 & 5.9 & 3.83 & 9 & 972 & 10 & 18 & 49 & 1864 & 11 & 4 & 91336 & 128.3 & 1.98 \\
        \hline
\end{tabular}

\vspace{0.1cm}
There are 5,046,621 data points with 114 of them on the Pareto frontier.
\label{tab:100k-rate0937}
\end{table*}


\end{document}